\documentclass[graybox]{svmult}

\usepackage{type1cm}        % activate if the above 3 fonts are
\usepackage{makeidx}         % allows index generation
\usepackage{graphicx}        % standard LaTeX graphics tool
\usepackage{multicol}        % used for the two-column index
\usepackage[bottom]{footmisc}% places footnotes at page bottom

\usepackage{newtxtext}       % 
\usepackage[varvw]{newtxmath}       % selects Times Roman as basic font

\newcommand{\dd}{\mathrm{d}}

\makeindex             % used for the subject index
\begin{document}

\title*{LVK constraints on PBHs from stochastic gravitational wave background searches}
\author{Alba Romero-Rodr\'iguez and Sachiko Kuroyanagi}
\institute{Alba Romero-Rodr\'iguez \at Theoretische Natuurkunde and IIHE/ELEM, Vrije Universiteit Brussel, \& The International Solvay Institutes, Pleinlaan 2, B-1050 Brussels, Belgium. 
\email{alba.romero-rodriguez@vub.be}
\and 
Sachiko Kuroyanagi \at Instituto de F\'isica Te\'orica UAM-CSIC, Universidad Aut\'onoma de Madrid, 28049 Madrid, Spain and Department of Physics, Nagoya University, Furo-cho Chikusa-ku, Nagoya 464-8602, Japan.  
\email{sachiko.kuroyanagi@csic.es}
}

\maketitle

\abstract{
Primordial black holes (PBHs) may have left an imprint in the form of a stochastic gravitational wave background (SGWB) throughout their evolution in the history of the Universe. This review highlights two types of SGWB: those generated by scalar curvature perturbations associated with PBH formation in the early Universe and those composed of ensembles of GWs emitted by PBH binaries. After describing detection methods and a brief introduction on Bayesian inference, we discuss current constraints imposed by LIGO-Virgo-KAGRA (LVK) observations through the non-detection of the SGWBs and discuss their physical implications.
}

\section{Introduction}
\label{sec:introduction}

A stochastic gravitational wave background (SGWB) offers an exciting opportunity to explore primordial black holes (PBHs). The SGWB is a random and persistent background of gravitational waves (GWs) coming from all directions. It is commonly referred to as `stochastic' because it can only be characterized statistically. The sources can be either cosmological or astrophysical. The former serves as a unique probe of the primordial Universe since it can propagate directly from the early epoch of inflation. The latter is formed by an ensemble of unresolved astrophysical events, and is also important for acquiring information on astrophysical source populations at high redshift, as well as for verifying the existence of PBH binaries. 

This review aims to discuss the implications for PBHs that can be obtained from SGWB observations. In particular, we review the constraints on PBH models derived from the nondetection in the most recent observation, namely Observation run 3 (O3) of the LIGO-Virgo-KAGRA (LVK) detector network. We focus on two different types of SGWBs: one arising from scalar curvature perturbations associated with PBH formation and the other generated from the collective GWs emitted by PBH binaries.

The first source is often termed the scalar-induced GW background (SIGWB). If there is a mechanism to amplify small-scale density perturbations in the early Universe, PBHs can form due to the collapse of extremely dense regions. In the scenario where primordial curvature fluctuations are amplified during inflation, PBHs form soon after the corresponding modes enter the horizon. Associated with this process, SGWBs are sourced by the second-order terms of scalar perturbations, which is predicted in the cosmological perturbation theory~\cite{Tomita:1975kj, Matarrese:1993zf,Ananda:2006af, Baumann:2007zm, Saito:2008jc, Domenech:2021ztg}. In this case, the peak scale is determined by the detailed mechanism of enhancing scalar perturbations during inflation, and it corresponds to the frequency of GWs, denoted by $f$,  as
\begin{equation}
f = \frac{c k}{2\pi} =25\left(\frac{k}{1.6 \times 10^{16}~\rm Mpc^{-1}}\right) {\rm Hz} 
\simeq 5.8 \times 10^{-9}\alpha^{1/2}\left(\frac{M_{\rm PBH}}{M_\odot}\right)^{-1/2} {\rm Hz} 
\,,
\label{eq:freq_SIGWB}
\end{equation}
where $k$ is the scale of the curvature perturbations, $M_\odot \simeq 2 \times 10^{33}$g denotes the solar mass, $\alpha\coloneqq M_{\rm PBH}/M_H$, typically of order unity, is used to account for the deviation between the PBH mass $M_{\rm PBH}$ and the horizon mass $M_H$~\footnote{Here, we have assumed the simplest scenario in which PBHs form with a fixed fraction of the horizon mass. However, to be precise, simulations of PBH formation show that the PBH mass depends on both the scale and amplitude of the perturbation from which it forms.}. GW experiments can probe scales much smaller than those accessible with current cosmological observations. Thus, they offer a unique opportunity to explore different epochs of inflation.

Second, the SGWB originates from the collective GWs emitted by PBH binaries. In this case, the SGWB consists of an ensemble of a large number of weak, independent, and unresolved GW sources~\cite{Phinney:2001di,Regimbau:2011rp, Cornish2015WhenIA}. Differences between primordial and astrophysical black holes typically emerge at high redshifts as PBHs form much earlier than astrophysical black holes. The SGWB provides a unique opportunity to probe such high redshifts and may offer hints to discriminate between astrophysical and primordial scenarios~\cite{Wang:2016ana, Mukherjee:2021ags, Mukherjee:2021itf, Bavera:2021wmw,Atal:2022zux}. Additionally, the PBH mass function can be very different from astrophysical ones, and this difference could be prominently reflected in the spectral shape~\cite{Braglia:2021wwa, Bagui:2021dqi}. This SGWB exhibits a peak at the merger frequency (redshifted), which is determined by the mass of PBHs. Assuming equal-mass binaries, the energy spectrum has a peak roughly at
\begin{equation}
f \simeq 8.3 \times 10^3 \left(\frac{M_{\rm PBH}}{M_\odot}\right)^{-1} {\rm Hz} \,.
\label{eq:freq_binary}
\end{equation}
The spectrum at low frequencies is attributed to the inspiral phase, which can be tested by the LVK detectors for solar mass PBHs.

As observed in Eqs.~\eqref{eq:freq_SIGWB} and \eqref{eq:freq_binary}, the two SGWB are generated by different mechanisms at different times and emerge in completely different frequency bands with different mass dependencies. Therefore, they can be investigated in multi-wavelength GW experiments~\cite{Wang:2019kaf}, including ground-based/space-borne interferometers, pulsar timing, etc. In this review, we focus on constraints obtained by the current LVK detectors, which explore two different PBH masses: assuming $f=25$Hz, we are probing PBHs with a mass of $5.0\times 10^{-20} M_\odot$ with SIGWB, while for PBH binaries, the corresponding redshifted PBH mass is $300 M_\odot$ (due to the low-frequency inspiral spectrum mentioned earlier, it can also access lower mass PBHs).

We aim to provide an overview of LVK SGWB constraints, hoping to serve as a useful guide for individuals interested in deriving constraints on their theoretical models from the data. The review is organized as follows: first, in Sec.~\ref{sec:analysis_techniques_LVK}, we briefly summarize the standard detection methods of SGWB, which involve cross-correlation between data from two or more interferometers, as well as the basics of Bayesian inference, which is often used to provide constraints on model parameters. Subsequently, in Sec.~\ref{sec:constraints_SIGWB}, we describe the current constraints on the SIGWB and provide their implications for PBH scenarios. In Sec.~\ref{sec:constraints_binaries}, we discuss the SGWB from PBH binaries and its constraints by the current LVK observations. We conclude in Sec.~\ref{sec:conclusions} describing future prospects.

\section{Analysis techniques for the SGWB search in the LIGO-Virgo-Kagra collaboration}
\label{sec:analysis_techniques_LVK}

\subsection{Basics and Assumptions}
\begin{figure}
    \centering
    \includegraphics[width=0.75\linewidth]{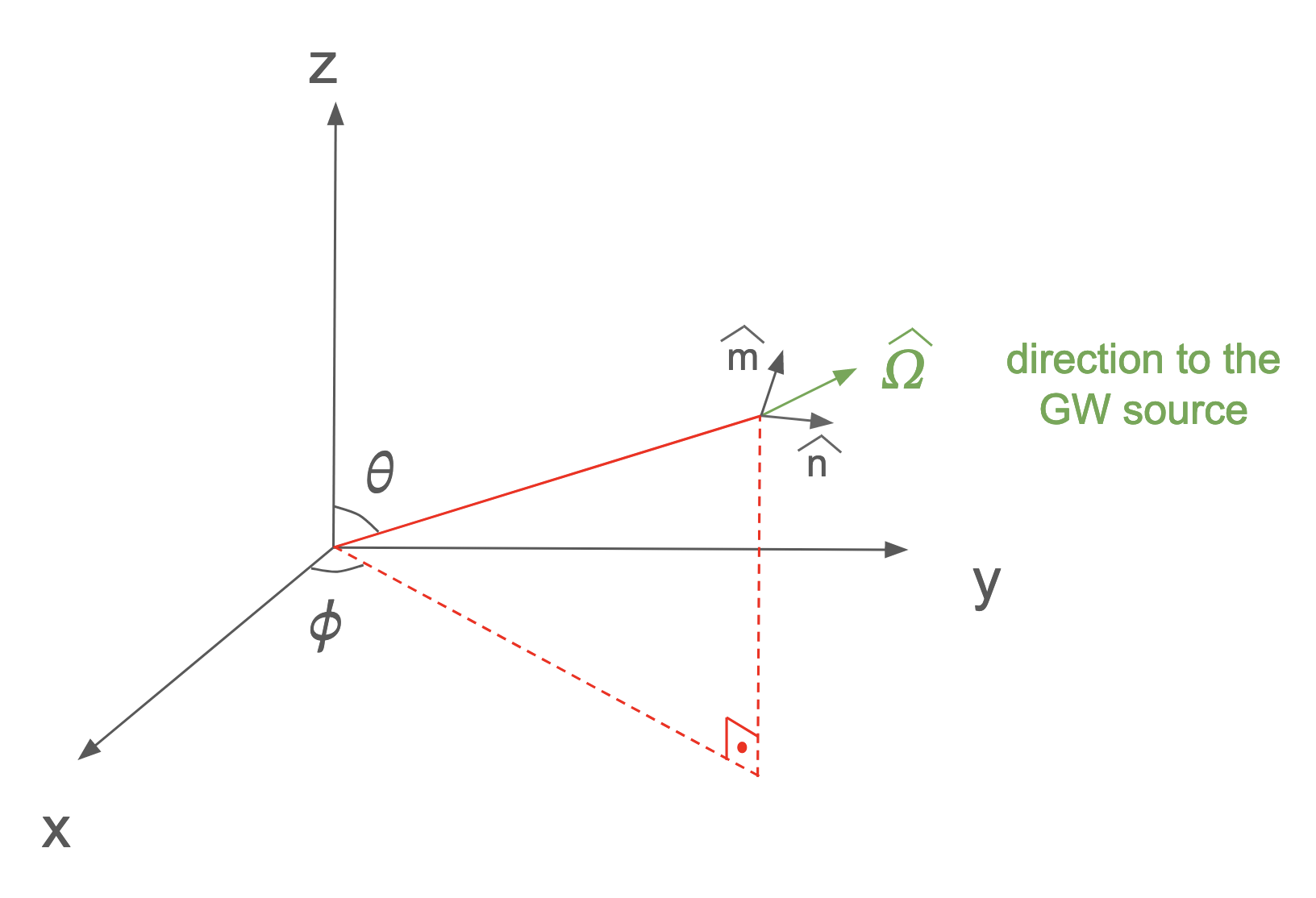}
    \caption{ The coordinate system consists of the unit vectors ${\hat{\Omega},\hat{m},\hat{n}}$, which define the polarization tensors $e_{ab}^ A (\hat{\Omega})$. The unit vector $\hat{\Omega}$ points in the direction of the GW propagation. The other two unit vectors lie perpendicular to $\hat{\Omega}$. }
    \label{fig:Pol_basis_unit_vectors}
\end{figure}

GWs can be described as a small tensor perturbation around the Minkowski background metric, i.e., $g_{ab} = \eta_{ab} + h_{ab}$ where $\eta_{ab} = {\rm diag}(-1,1,1,1)$ and the tensor perturbations $h_{ab}$ satisfy the transverse traceless conditions. They are often expressed as a superposition of plane waves by decomposing into their Fourier modes using the two independent polarization states
\begin{equation}
 h_{ab} (t,\textbf{x}) = \sum_{A = +,\times} \int_{-\infty}^\infty \dd f\int_{S^2} \dd\hat{\Omega} \,
 h_A(f,\hat{\Omega}) e^{2\pi if(t-\hat{\Omega}\cdot \textbf{x}/c)}e_{ab}^A(\hat{\Omega}) \,, 
 \label{eq:h_ab}
\end{equation}
where $f$ is the GW frequency. The unit vector $\hat{\Omega}$ represents the direction of propagation of the wave and is related to the wave vector as $\textbf{k}=2\pi f \hat{\Omega}/c$. Each plane wave is characterized by the Fourier amplitude $h_A(f,\hat{\Omega})$ and the phase $2\pi if(t-\hat{\Omega}\cdot \textbf{x}/c)$. The metric perturbations $ h_{ab}(t,\textbf{x})$ are real numbers, so the plane wave components satisfy $ h_A(-f,\hat{\Omega})=h_A^*(f,\hat{\Omega})$. The polarization tensors $e_{ab}^A (\hat{\Omega})$ denote the two independent polarizations of GWs. Although the choice of polarization basis is not unique and may be chosen based on the specific problem, here we use the most common basis to define the plus and cross polarizations ($A = +, \times$)
\begin{eqnarray}
e_{ab}^ + (\hat{\Omega}) &=& \hat{m}_a\hat{m}_b-\hat{n}_a\hat{n}_b \,, \label{eq:plus_pol_tensor} \\
e_{ab}^ \times (\hat{\Omega}) &=& \hat{m}_a\hat{n}_b+\hat{n}_a\hat{m}_b \,, \label{eq:cross_pol_tensor}
\end{eqnarray}
where $\hat{\Omega}$, $\hat{m}$ and $\hat{n}$ are the unit vectors (see Fig.~\ref{fig:Pol_basis_unit_vectors}), defined as
\begin{eqnarray}
\hat{\Omega} &=& \cos{\phi} \sin{\theta} ~\hat{x}+\sin{\phi}\sin{\theta}~\hat{y}+\cos{\theta}~\hat{z} \,,
    \label{eq:unit_vec_1}\\
\hat{m} &=& \sin{\phi}~\hat{x}-\cos{\phi}~\hat{y} \,,
    \label{eq:unit_vec_2}\\
\hat{n} &=& \cos{\phi}\cos{\theta}~\hat{x}+\sin{\phi}\cos{\theta}~\hat{y}-\sin{\theta}~\hat{z} \,.
    \label{eq:unit_vec_3}
\end{eqnarray}
They satisfy $e^A_{ab}(\hat{\Omega})e^{A', ab} (\hat{\Omega}) = 2\delta^{AA'}$.

The metric perturbations  $h_{ab}$ and the Fourier amplitudes $h_{A}(f,\hat{\Omega})$ are random variables. Hence, the stochastic background is characterized statistically. Due to the central limit theorem, the SGWB is often assumed to follow a Gaussian distribution. This assumption arises from the consideration that, in most cases of cosmological backgrounds, a very large number of Hubble patches in which GWs are produced are observed across the resolution area of the detector. For astrophysical backgrounds, it holds if it comprises a sufficiently large number of overlapped independent sources. The assumption of Gaussianity simplifies the analysis because the entire probability density function can be characterized by the quadratic expectation value, i.e., $\langle  h_A^* (f,\hat{\Omega})h_{\rm A'}(f',\hat{\Omega}')\rangle$. Note that a SGWB has a zero mean, indicating $\langle h_A (f,\hat{\Omega})\rangle = 0$. 

\begin{backgroundinformation}{Background Information - Random variables}
The statistical properties of a random variable $X$ are determined by its probability distribution $p_{\rm X}(x)$. The $n$-th order moments of the distribution are defined by $\langle X^n \rangle \coloneqq \int ^\infty_{-\infty} \dd x x^n p_{\rm X}(x)$. The first moment, $\langle  X \rangle$, is the mean value of $X$ and is usually referred to as $\mu$. The second moment, $\langle  X^2 \rangle$, is related to the variance of the distribution as $\sigma^2 = \langle X^2 \rangle - \langle X \rangle ^2$. 

If $X$ is a Gaussian random variable, then its probability distribution can be expressed as 
\begin{equation}
    p_{\rm X}(x) = \frac{1}{\sqrt{2\pi}\sigma}e^{-\frac{(x-\mu)^2}{2\sigma^2}},
\end{equation}
which implies that $X$ is entirely described by its first two momenta. In this case, all the higher-order moments are either zero or can be expressed as a sum of products of the first two momenta (Isserlis' theorem~\cite{isserlis1918}). 

\end{backgroundinformation}

In this review, we only focus on a Gaussian, stationary, unpolarized, and isotropic SGWB. This assumption is valid in the case of the SIGWB. The SIGWB signal is intrinsically non-Gaussian, unlike other cosmological sources, because its source is second-order in the curvature perturbation. However, the propagation effects of the GWs across the perturbed Universe suppress the level of non-Gaussianity to an unobservable level~\cite{Bartolo:2018evs,Bartolo:2018rku}. In the case of the SGWB from PBH binaries (and in general astrophysical backgrounds), it can deviate from this assumption and the SGWB can exhibit non-Gaussianity~\cite{Braglia:2022icu}, anisotropy~\cite{Wang:2021djr}, and polarizations~\cite{ValbusaDallArmi:2023ydl}. In fact, these additional statistics contain rich information on the black hole population and help differentiate between different PBH models.

Another major assumption that we make in this section concerns Gaussian, stationary, and uncorrelated detector noise. Real-world detectors experience numerous non-Gaussian transient noises, often termed glitches, along with non-stationary noise characterized by slow variations in detector sensitivity. The stochastic analysis in the LVK~\cite{PhysRevD.91.022003,PhysRevLett.118.121101,PhysRevD.100.061101,PhysRevD.104.022004} is conducted to ensure that the Gaussianity and stationarity of noise are consistently maintained through preprocessing the strain data with the procedure known as {\it gating} and implementing the so-called {\it delta-sigma cut}. During the gating process, loud glitches are eliminated by applying an inverse Tukey window to instances where the data exhibits a significant excess~\cite{gating}. Residual non-Gaussian noises and non-stationarities are also addressed through the delta-sigma cut by monitoring the standard deviation of each time-series segment and removing all times where segments vary too much compared to adjacent segments. See Refs.~\cite{Renzini_2023,vanRemortel:2022fkb} for details of data conditioning.

Furthermore, specific sources of correlated noise exist, deviating from our idealized framework. Noteworthy are narrow-frequency noise artifacts such as correlations between the electronics mains frequency (e.g., 60Hz in LIGO and 50Hz in Virgo) and data sampling referenced to GPS clocks~\cite{LSC:2018vzm}. These known noise lines are typically {\it notched out}, meaning that the values of the spectra at the affected frequency bins are removed from the analysis. Of particular concern are Schumann resonances, electromagnetic phenomena originating from lightning strikes that persist within the Earth's ionosphere, resulting in coherent oscillations in magnetometer readings at ground-based GW detectors. These resonances, primarily below 50Hz, present a significant challenge due to their potential coupling with mirror suspension systems, electric cables, and electronics. A noise budget for magnetic correlations was constructed in the observing runs of LIGO-Virgo data~\cite{PhysRevD.87.123009,PhysRevD.90.023013,PhysRevD.104.022004}. Despite the current findings indicating no significant evidence for correlated magnetic noise, future sensitivities demand vigilant consideration of these resonances~\cite{PhysRevD.104.122006}. One of the solutions to this challenge could be GW geodesy~\cite{Callister_2018,PhysRevD.105.082001}, a new robust method that enables us to discern between potential correlated noises and genuine SGWB signals by requiring true observations to be consistent with the known geometry of the detector network.

\subsection{Gaussian, stationary, unpolarized, and isotropic SGWB}
\label{sec:isotropic_GWB}

The SGWB, being Gaussian with zero mean, isotropic, unpolarized, and stationary allows us to write the 2-point correlation function as
\begin{equation}
\langle  h_A^* (f,\hat{\Omega})h_{A'} (f',\hat{\Omega}')\rangle
= \frac{1}{16\pi}S_h(f) \delta_{AA'}\delta(f-f')\delta^2(\hat{\Omega},\hat{\Omega'}) \,,
    \label{eq:2_pt_corr_func}
\end{equation}
where $S_h(f)$ is called the strain power spectral density (PSD) and satisfies $S_h(f)=S_h(-f)$. The SGWB is often characterized by the dimensionless energy density parameter
\begin{equation}
    \Omega_{\rm GW} (f) = \frac{1}{\rho_{c,0}}\frac{\dd\rho_{ \rm GW}}{\dd( \ln f)} \,,
    \label{eq:Omega_GW}
\end{equation}
where $\rho_{c,0}=3 H_0^2c^2/(8\pi G)$ is the critical energy density of the Universe today with $H_0$ being the Hubble parameter at present time. The energy density of GWs, $\rho_{\rm GW}$, is expressed as
\begin{equation}
    \rho_{\rm GW} = \frac{ c^2}{32 \pi  G} \langle  \dot{h}_{ab}(t,\textbf{x})\dot{h}^{ab}(t,\textbf{x})\rangle \,.
    \label{eq:rho_GW}
\end{equation}
By substituting (\ref{eq:h_ab}), we obtain
\begin{eqnarray}
    \rho_{\rm GW}  & = &  \frac{c^2}{32\pi G}\sum_{A,A'=+, \times} 
    \int_{-\infty}^{\infty}\dd f \int_{-\infty}^{\infty}\dd f'\int_{S^2} \dd\hat{\Omega} \int_{S^2} \dd\hat{\Omega}'\langle h_A^*(f,\hat{\Omega})h_{A'}(f,\hat{\Omega})\rangle \nonumber \\ 
    & & \times \,\, 4\pi^2ff' e^{-2\pi if(t-\hat{\Omega}\cdot\textbf{x}/c)}e^{2\pi if'(t-\hat{\Omega}'\cdot\textbf{x}/c)}e_{ab}^{A}(\hat{\Omega})e^{A',ab}(\hat{\Omega}')
    \nonumber \\
    & = &  \frac{c^2}{128 G}\sum_{A,A'=+, \times} \int_{-\infty}^{\infty}\dd f \int_{-\infty}^{\infty}\dd f'\int_{S^2} \dd\hat{\Omega} \int_{S^2} \dd\hat{\Omega}'S_h(f) \delta_{AA'}\delta(f-f')\delta^2(\hat{\Omega},\hat{\Omega'}) \nonumber \\ 
    & &  \times \,\, ff' e^{-2\pi if(t-\hat{\Omega}\cdot\textbf{x}/c)}e^{2\pi if'(t-\hat{\Omega}'\cdot\textbf{x}/c)}e_{ab}^{A}(\hat{\Omega})e^{A',ab}(\hat{\Omega}')
    \nonumber \\
    & = & \frac{\pi c^2}{8 G}\int_{-\infty}^\infty \dd f \, f^2 S_h(f) =  \frac{\pi c^2}{4 G}\int_0^\infty \dd f \, f^2 S_h(f)\,.
\end{eqnarray}
In the third step, under the assumption of an unpolarized background, we have used $ \sum_{+,\times} e_{ab}^A(\hat{\Omega})e^{A,ab}(\hat{\Omega})=4$. Furthermore, assuming an isotropic background, the solid angle integral yields a factor of $4\pi$. In the last step, we have replaced the frequency integration range as $\int_{-\infty}^\infty \dd f = 2\int_0^\infty \dd f$. Substituting this result into Eq.~\eqref{eq:Omega_GW}, we obtain\footnote{Note that some literature (e.g. Ref.~\cite{Maggiore:1999vm}) has a factor of 2 difference in the definition of Eq.~\eqref{eq:2_pt_corr_func}, resulting in a factor of 2 difference in $\Omega_{\rm GW}$. We follow the definition in Refs.~\cite{Allen_Romano_99,Romano_Cornish_2017}, which is commonly used in LVK stochastic papers.}
\begin{equation}
\Omega_{\rm GW} (f) = \frac{2\pi^2}{3H_0^2}f^3 S_h(f) \,.
\label{eq:OmegaGW_Sh}
\end{equation}
This equation implies that, because of the $f^3$ dependence, given two detectors with the same strain sensitivity, the one at lower frequencies is sensitive to lower values of $\Omega_{\rm GW}(f)$. Consequently, for instance, in the case of LVK O3 analysis, the most sensitive frequency band for stochastic search lies in the relatively low-frequency range of $30-40$Hz, while the noise floor is observed above $100$Hz when sensitivity is plotted in terms of the usual strain noise amplitude.

\begin{figure}[!htbp]
    \centering
    \includegraphics[width=\linewidth]{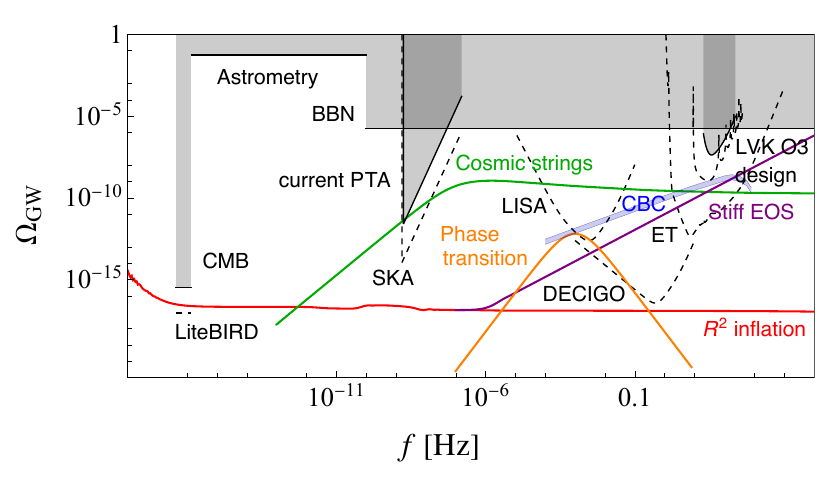}
   \caption{SGWB spectra for several cosmological models (colored solid curves) are compared to current upper bounds (shaded in gray) and expected sensitivities of future experiments (black dashed lines). As examples of cosmological sources, we show GW signals from inflation (assuming $R^2$ inflation~\cite{Starobinsky:1980te}) and its modification by a stiff Equation of State~\cite{Giovannini:1998bp}. Note that the amplitude and spectral behavior of these signals heavily depend on the model parameters, and the spectra shown here represent just one example of this variability. We also depict the GW spectrum of cosmic string loops~\cite{Damour:2000wa,Damour:2001bk} (assuming tension $G\mu = 10^{-12}$) and the electroweak phase transition~\cite{Kosowsky:1992vn}. Current constraints include the Advanced-LIGO O3 upper bound~\cite{PhysRevD.104.022004}, constraints based on big bang nucleosynthesis (BBN) and cosmic microwave background (CMB) observations~\cite{Yeh:2022heq}, pulsar timing array~\cite{InternationalPulsarTimingArray:2023mzf}, astrometric measurement by Gaia satellite~\cite{Jaraba:2023djs}, and CMB temperature and polarization observation~\cite{BICEP:2021xfz}. Future expected sensitivities include the final sensitivity of Advanced-LIGO~\cite{PhysRevLett.116.131102}, ET~\cite{Abbott_2017}, DECIGO~\cite{Kawamura:2020pcg}, LISA~\cite{Colpi:2024xhw}, SKA~\cite{Dewdney:2009tmd}, and LiteBIRD~\cite{LiteBIRD:2020khw}. We assume $3$ years of observation for interferometer experiments and $20$ years for the SKA. The blue-shaded band indicates the expected amplitude of the SGWB due to the cosmic population of compact binary coalescences (CBCs), based on the observed individual events in the O3 catalog~\cite{KAGRA:2021duu}. Note that the expected spectral amplitude is extrapolated to the LISA frequency band assuming the $f^{2/3}$ dependence of the inspiral phase~\cite{Babak:2023lro, Lehoucq:2023zlt}. However, the lower frequencies could be modified by the effects of eccentricity and precession at the time of binary formation~\cite{DOrazio:2018jnv,Zhao:2020iew}.
 }
\label{fig:Comparison_models}
\end{figure}

Now that the energy density spectrum has been introduced, Figure~\ref{fig:Comparison_models} displays examples of the spectrum of the SGWB for different cosmological models, alongside the current best upper bounds and future expected GW detector sensitivities. Many generation mechanisms predict the existence of a SGWB across a wide range of frequencies, emphasizing the importance of multi-band GW observations in the future. Current ground-based detectors have limited sensitivity to probe possible early Universe signals. However, future detectors such as the Laser Interferometer Space Antenna (LISA) and third-generation detectors like the Einstein Telescope (ET) could investigate specific sources such as inflation, cosmic strings, and phase transitions.

\subsection{SGWB detection method: cross-correlation technique}
\label{sec:detection_isotropic_GWB}

In this subsection, we outline the framework for the standard detection method. Given the stochastic nature of a SGWB, distinguishing between a signal and local detector noise poses a significant challenge. Thus, SGWB searches often involve cross-correlating data from pairs of interferometers\footnote{Note that LISA (and possibly ET, whose design is still under consultation) uses the so-called Time Delay Interferometry (TDI) variables for data processing. In this method, six laser links are transformed into a null channel containing mainly noise and two orthogonal signal channels, where the noise is suppressed. In this case, two orthogonal signal channels are not correlated (as they effectively correspond to two L-shape detectors; one of which is rotated by $\pi/4$ radians with respect to the other~\cite{Cutler:1997ta}), so the auto-correlation of the two signal channels is used to search for a SGWB~\cite{Smith:2019wny}.}~\cite{1987MNRAS.227..933M,Christensen:1992wi,Flanagan:1993ix}. For a detailed review, we refer the reader to Ref.~\cite{Allen_Romano_99,Romano_Cornish_2017}. 
 
Let us define the output of a detector $s_{\rm i}(t)$ as the sum of the detector's noise $n_{\rm i}(t)$ and a GW signal $h_{\rm i}(t)$, where the label i denotes each detector,
\begin{equation}
\begin{split}
     s_1(t)=&  n_1(t)+h_1(t),\\
     s_2(t)=&  n_2(t)+h_2(t)\,.
\end{split}
    \label{eq:detector_output}
\end{equation}
The basic concept is that, when we correlate the outputs of the two detectors and integrate them over time, the noise terms cancel out because they are zero-mean random variables $\langle n_{\rm i}(t) \rangle = 0$ and ideally have no correlation with each other $\langle n_{\rm i}(t)n_{\rm j}(t) \rangle \propto \delta_{\rm ij}$. As a simple example, if we have two co-located and co-aligned detectors, so that $h_1(t)=h_2(t)=h(t)$, we can extract the signal in the following way:
\begin{eqnarray}
    \int_{-T/2}^{T/2} \dd t ~ s_1(t) s_2(t)
    &=& \int_{-T/2}^{T/2} \dd t \left(n_1(t) n_2(t) + h_1(t) n_2(t) + n_1(t) h_2(t) + h(t)^2\right)
    \nonumber \\
    &\simeq& \int_{-T/2}^{T/2} \dd t ~ h(t)^2 = T \langle h(t)^2 \rangle \,.
\end{eqnarray}

In reality, the situation is more complex. The LVK analysis employs the optimal filtering method, specifically designed to detect a particular spectral shape of a SGWB while accounting for the locations of the detectors. It is equivalent to constructing the cross-correlation estimator for an observation period $T$ using the output of the two detectors $s_1$ and $s_2$ as
\begin{equation}
 Y \coloneqq \int_{-T/2}^{T/2}\dd t \int_{-T/2}^{T/2}\dd t' s_1(t)s_2(t)Q(t,t').
    \label{eq:cross_correlation_estimator}
\end{equation}
The function $Q(t,t')$ is known as a filter function, chosen to maximize the signal-to-noise ratio (SNR) for given detector locations and the spectral shape of the GW signal and detector noise, as we will see shortly. Since we have to determine the spectral shape of the SGWB, the Fourier transform of this function is sometimes referred to as a ``template''. As described in the following section, to constrain a particular model, one has to prepare a set of spectral shapes predicted by theory (a ``template bank'' in the terminology of matched filtering search) and find the most probable template by iteratively calculating the estimator of Eq.~\eqref{eq:cross_correlation_estimator} for different templates. 

Noise and SGWB are assumed to be stationary during the observation time $T$, so the optimal choice of the filter function depends only on the time difference $\Delta t = t - t'$ and $Q(t,t')$ can be written as $Q(t - t')$. Note that the cross-correlation is maximum when two detectors are co-located and co-aligned, and the filter function has a peak at $t=t'$. It falls off rapidly to zero when $\Delta t$ is much larger than the light travel time between the two sites. Using this fact, one can extend the integration range of $t'$ to [$-\infty, \infty$] and perform the Fourier transformation as
\begin{equation}
     Y =   \int_{-\infty}^{\infty} \dd f \int_{-\infty}^{\infty} \dd f'
     \delta_{\rm T}(f-f') s_1^*(f)s_2(f')Q(f') \,,
    \label{eq:cross_correlation_estimator_2}
\end{equation}
where $\delta_{\rm T}(f-f')$ is a finite time approximation to the Dirac delta function, 
\begin{equation}
     \delta_{\rm T}(f-f')=\int_{-T/2}^{T/2}\dd t 
     ~ e^{-2\pi i (f-f')t}=\frac{\sin{(\pi T (f-f'))}}{\pi (f-f')}.
    \label{eq:finite_time_delta}
\end{equation}
For $f=f'$, we have $\delta_{\rm T} (0)\rightarrow T$, and we get the usual Dirac delta function for $ T\rightarrow \infty$. 

The expectation value of $Y$, denoted by $\langle Y \rangle$, has an associated variance given by $\sigma_{\rm Y}^2 = \langle Y^2 \rangle - \langle Y \rangle^2$. Thus, we define the signal-to-noise ratio as 
\begin{equation}
{\rm SNR}^2\coloneqq \frac{\langle Y\rangle^2}{\sigma_{Y}^2} \,.  
\end{equation}
To calculate $\langle Y\rangle$, we need to evaluate $\langle s_1^*(f)s_2(f') \rangle$. Using Eq.~(\ref{eq:detector_output}) and assuming uncorrelated random noises, this expectation value can be approximated as 
\begin{eqnarray}
     \langle s_1^*(f)s_2(f') \rangle 
     &=& \langle n_1^*(f)n_2(f') + h_1^*(f)n_2(f') + n_1^*(f)h_2(f') + h_1^*(f)h_2(f') \rangle \nonumber\\
     &\simeq & \langle h_1^*(f)h_2(f') \rangle \,.
    \label{eq:expec_value_s1_s2}
\end{eqnarray}

Now let us remove the assumption of co-located and co-aligned detectors. The GW signal at detector i can be expressed as the contraction between the metric and the detector $ h_{\rm i} (t) = h_{ab} (t, \textbf{x}_{\rm i}) d^{\rm ab}(t, \textbf{x}_{\rm i})$, where $\textbf{x}_{\rm i}$ is the location of the detector. The vector $ d^{\rm ab}(t, \textbf{x}_{\rm i})$ denotes the response of the detector i located at $\textbf{x}_{\rm i}$ and time $t$. When the round-trip light travel time in the detector arms is short compared to the period of the GWs (which is the case for the LVK detectors), it can be expressed in terms of the unit vectors along the two laser arms of the detector, $\hat{X}(t,\textbf{x}_{\rm i})$ and $\hat{Y}(t,\textbf{x}_{\rm i})$,
\begin{equation}
 d^{\rm ab}(t, \textbf{x}_{\rm i})\simeq \frac{1}{2}\Big(\hat{X}^{\rm a}(t,\textbf{x}_{\rm i})\hat{X}^{\rm b}(t,\textbf{x}_{\rm i})-\hat{Y}^{\rm a}(t,\textbf{x}_{\rm i})\hat{Y}^{\rm b}(t,\textbf{x}_{\rm i})\Big).
    \label{eq:detector_response}
\end{equation}
Then the Fourier amplitude is expressed as
\begin{equation}
 h_{\rm i}(f;t) = \sum_{A=+,\times} \int_{S^2} \dd\hat{\Omega} \,
 h_A (f,\hat{\Omega}) e^{-2\pi i f \hat{\Omega}\cdot \textbf{x}_{\rm i} /c} e^{\rm A}_{ab}(\hat{\Omega})d^{\rm ab}(t,\textbf{x}_{\rm i})\,.
    \label{eq:h_i_freq_domain}
\end{equation}
In the equation, we see the contraction of the polarization tensor with the detector response tensor at location $\textbf{x}_{\rm i}$ and time $t$, which is often defined as $F_{\rm i} ^ {\rm A} (\hat{\Omega},t) \coloneqq e^{\rm A}_{ab}(\hat{\Omega})d^{\rm ab}(t,\textbf{x}_{\rm i})$ and called the \textit{detector pattern functions}. Here the $t$ index is kept to emphasize the fact that the analysis is done per segment of time. In reality, the direction of the detectors changes due to the Earth's motion, and this must be carefully considered when the SGWB is anisotropic. In the following, we omit this dependence by focusing on the isotropic case, where detectors always receive the same contributions from the SGWB regardless of changes in the direction of the detector over time.

Then, Eq.~(\ref{eq:expec_value_s1_s2}) is now expressed as 
\begin{eqnarray}
    && \langle s_1^*(f)s_2(f') \rangle \nonumber\\
    & \simeq &  \sum_{A,A'=+,\times} \int_{S^2} \dd\hat{\Omega} \int_{S^2} \dd\hat{\Omega}'\langle h_A^* (f,\hat{\Omega})h_{\rm A'} (f',\hat{\Omega}')\rangle \, e^{2\pi i f \hat{\Omega} \cdot \textbf{x}_1/c} e^{-2\pi i f \hat{\Omega}' \cdot \textbf{x}_2/c} 
    F_1^A (\hat{\Omega})F_2^{A'} (\hat{\Omega}') \nonumber \\
    &=&  \frac{3H_0^2}{32\pi^3 f^3} \Omega_{\rm GW}(f)\delta(f-f')\sum_{A,A'=+,\times} \int_{S^2} \dd\hat{\Omega} \, e^{2\pi i f \hat{\Omega} \cdot\Delta \textbf{x}/c} F_1^A (\hat{\Omega})F_2^{A'} (\hat{\Omega}) \,,
    \label{eq:expec_value_s1_s2_2}
\end{eqnarray}
where we have defined $\Delta \textbf{x} \coloneqq \textbf{x}_1 -\textbf{x}_2$, and have used Eqs.~\eqref{eq:2_pt_corr_func} and \eqref{eq:OmegaGW_Sh} in the second step.
Then the expectation value of the estimator $Y$ can be expressed as
\begin{equation}
     \langle Y \rangle = \frac{3H_0^2}{20\pi ^2} T 
     \int_{-\infty}^{\infty} \dd f \frac{\Omega_{\rm GW}(|f|)}{|f|^3}\gamma_{12}(|f|)Q(f) \,,
    \label{eq:cross_correlation_estimator_average_3}
\end{equation}
where $\gamma_{12}(f)$ is a purely geometrical quantity known as the {\it overlap reduction function} (ORF), given by
\begin{equation}
 \gamma_{12}(f) \coloneqq \frac{5}{8\pi}\sum_{A,A'=+,\times} \int_{S^2} \dd\hat{\Omega}e^{2\pi i f \hat{\Omega} \cdot\Delta \textbf{x}/c} F_1^A (\hat{\Omega})F_2^{A'} (\hat{\Omega}).
    \label{eq:ORF}
\end{equation}
The pre-factor $5/(8\pi)$ is chosen so that $\gamma_{12}(f)=1$ for co-aligned and co-located detectors. 

The next step consists of calculating the variance $\sigma_Y^2$.  Assuming the weak-signal limit, i.e. the detector noise is much larger than the SGWB signal $h_{\rm i}(t)\ll n_{\rm i}(t)$, we can approximate $\sigma_Y^2 = \langle Y^2 \rangle - \langle Y \rangle^2 \simeq  \langle Y^2 \rangle$.\footnote{See Ref.~\cite{Kudoh:2005as}, for the case of strong-signal regime, where we can no longer neglect the $\langle Y \rangle^2$ term.}  For a Gaussian random variable, all cumulants vanish aside from the 2-point cumulants. Thus, using the assumption of uncorrelated random noises, the 4-point correlation function of the noise is reduced to
\begin{eqnarray}
    &&\langle n_1^*(f)n_2(f')n_1^*(f'')n_2(f''')\rangle 
    \nonumber \\
    &&= \langle n_1^*(f)n_2(f')\rangle \langle n_1^*(f'')n_2(f''')\rangle 
    + \langle n_1^*(f)n_1^*(f'')\rangle \langle n_2(f')n_2(f''')\rangle 
    \nonumber \\
    && \hspace{1.5cm} 
    + ~ \langle n_1^*(f)n_2(f''')\rangle \langle n_2(f')n_1^*(f'')\rangle
    \nonumber \\
    &&\simeq \langle n_1^*(f)n_1^*(f'')\rangle \langle n_2(f')n_2(f''')\rangle \,.
\end{eqnarray}
The time-series noise $n_{\rm i}(t)$ has real numbers and its Fourier transform satisfies $n_{\rm i}^*(f)=n_{\rm i}(-f)$. The same holds for the filter function. Using this, we obtain
\begin{eqnarray}
     \sigma_Y^2 
     &\simeq& \int_{-\infty}^{\infty} \dd f \int_{-\infty}^{\infty} \dd f' \int_{-\infty}^{\infty} \dd f'' \int_{-\infty}^{\infty} \dd f''' \langle n_1^*(f)n_1(-f'')\rangle \langle n_2^*(-f')n_2(f''')\rangle \nonumber\\
     && \hspace{3.5cm} \times \, \delta_{\rm T}(f-f')\delta_{\rm T}(f''-f''')Q(f')Q(f''') \nonumber\\
     & = &  \frac{1}{4}\int_{-\infty}^{\infty} \dd f \int_{-\infty}^{\infty} \dd f' P_1(|f|)P_2(|f'|)\delta_{\rm T}^2(f-f')Q(f)Q^*(f')  \nonumber\\ 
    & = &  \frac{T}{4} \int_{-\infty}^{\infty} \dd f P_1(|f|)P_2(|f|) |Q(f)|^2 \,,
    \label{eq:sigma_Y_2}
\end{eqnarray}
where, in the second step, we have used the definition of the one-sided noise power spectrum\footnote{Frequencies are positive, leading to a factor of $1/2$ in the definition of the noise power spectrum. The absolute value over $f$ is used to emphasize $f\in \mathcal{R}^+$.}
\begin{equation}
\langle n_{\rm i}^*(f)n_{\rm i}(f')\rangle = \frac{1}{2}\delta(f-f')P_{\rm i}(|f|) \,.
\end{equation}
In the last equality, $\delta_{\rm T}(0)=T$ is used. 

\begin{figure}[!htbp]
\centering
    \includegraphics[width=0.7\linewidth]{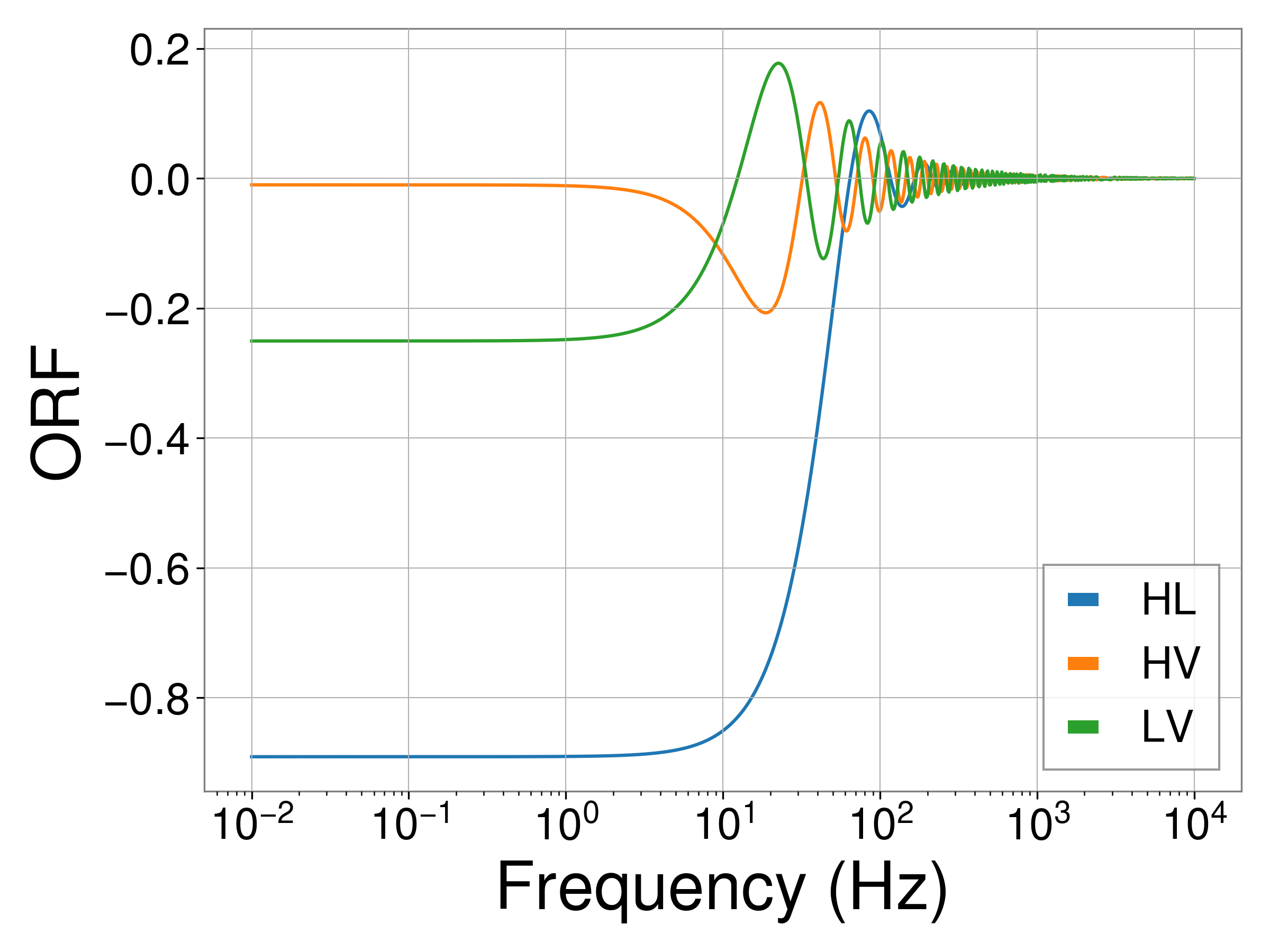}
    \caption{Overlap reduction function (ORF) between the three LIGO-Virgo interferometers, where `H' indicates Hanford, `L' denotes Livingston, and `V' represents Virgo. The curves are calculated by substituting actual relative positions and orientations of detector pairs (see e.g., Ref.~\cite{Seto:2008sr} for parameter values). The ORF starts to oscillate and decay rapidly above the characteristic frequency $f_c = c/(2\Delta {\bf x})$. This implies that nearby detectors in a pair lose less sensitivity at low frequencies and are more sensitive to a SGWB.
    The cases in which the ORF is negative as $f\rightarrow 0$ are due to the fact that the two interferometers considered for the calculation are rotated by $\pi/2$ radians with respect to each other.}
    \label{fig:Romano_paper_ORF}
\end{figure}

Now we define an inner product between functions $ A(f)$ and $ B(f)$ as
\begin{equation}
 (A,B)\coloneqq\int_{-\infty}^{\infty}\dd fA^*(f)B(f)P_1(|f|)P_2(|f|).
    \label{eq:Inner_product}
\end{equation}
Then the expectation value of the cross-correlation estimator in Eq.~(\ref{eq:cross_correlation_estimator_average_3}) can be written as
\begin{equation}
 \langle Y \rangle = \frac{3H_0^2T}{20\pi^2}\Big(Q(|f|),\frac{\Omega_{\rm GW}(|f|)\gamma_{12}(|f|)}{f^3P_1(|f|)P_2(|f|)}\Big).
    \label{eq:cross_correlation_estimator_average_2}
\end{equation}
Similarly, the variance of $Y$, given in  Eq.~(\ref{eq:sigma_Y_2}), can be written as 
\begin{equation}
 \sigma_{\rm Y}^2 = \frac{T}{4}(Q(|f|),Q(|f|)).
    \label{eq:var_Y_2}
\end{equation}
Therefore, the SNR is described as  
\begin{equation}
{\rm SNR}^2 \coloneqq \frac{\langle Y\rangle^2}{\sigma_{\rm Y}^2}=\left(\frac{3H_0^2}{10\pi^2}\right)^2 T \frac{\left(Q(|f|),\frac{\Omega_{\rm GW}(|f|)\gamma_{12}(|f|)}{f^3P_1(|f|)P_2(|f|)}\right)^2}{(Q(|f|),Q(|f|))}.
    \label{eq:SNR}
\end{equation}
The filter function $Q(|f|)$ is chosen such that the SNR is maximized. This happens when $ Q(|f|)$ is
\begin{equation}
 Q(|f|) = \lambda\frac{\Omega_{\rm GW}(|f|)\gamma_{12}(|f|)}{f^3P_1(|f|)P_2(|f|)},
    \label{eq:Q_filter}
\end{equation}
where $\lambda$ is a %scaling factor. 
overall normalization constant.
Substituting Eq.~(\ref{eq:Q_filter}) into Eq.~(\ref{eq:SNR}) results in
\begin{equation}
{\rm SNR} = \frac{3H_0^2\sqrt{T}}{10\pi^2}  \Bigg(\int_{-\infty}^{\infty}\dd f\frac{\Omega_{\rm GW}^2(|f|)\gamma_{12}^2(|f|)}{|f|^6P_1(|f|)P_2(|f|)}\Bigg)^{1/2}.
    \label{eq:SNR_2}
\end{equation}
Major implications deriving from this equation are explained as follows:
\begin{itemize}
    \item The SNR increases proportionally to $\sqrt{T}$.
    \item The GW signal amplitude $\Omega_{\rm GW}(|f|)$ in the numerator is multiplied by the ORF, introduced in Eq.~(\ref{eq:ORF}). It quantifies the reduction in sensitivity of the cross-correlation due to the response of the detectors as well as their separation and orientation. In Fig.~\ref{fig:Romano_paper_ORF}, we plot the ORF between the LIGO and Virgo interferometers. 
    \item The noise power spectra $P_{\rm i}(|f|)$ in the denominator represent the detector noise. A smaller $P_{\rm i}(|f|)$ leads to better detector sensitivity, resulting in an increased SNR. This naturally suppresses noisy frequencies from the analysis. 
    \item The filter function to maximize the SNR includes the spectral shape $\Omega_{\rm GW}(|f|)$, which is not known a priori. Therefore, we need to prepare a set of filters (templates) and apply them iteratively to find the maximum SNR.
\end{itemize}

\begin{backgroundinformation}{Background information}
In the LVK stochastic analysis, the spectrum is usually modeled using a power law form
\begin{equation}
    \Omega_{\rm GW} (f) = \Omega_\alpha \left(\frac{f}{f_{\rm ref}}\right)^{\alpha},
\label{eq:Omega_GW_assumption}  
\end{equation}
where $\alpha$ denotes the spectral index, $f_{\rm ref}$ is the reference frequency, and $\Omega_\alpha$ represents the amplitude of the SGWB at $f_{\rm ref}$. Typically, $f_{\rm ref}$ is set to $25$Hz, as this frequency roughly corresponds to the peak sensitivity of the LIGO-Virgo network.

The typical values of the spectral index $\alpha$ are 0, 2/3, or 3.  For example, $\alpha = 0$ characterizes cosmic strings at the high-frequency end~\cite{PhysRevLett.98.111101} and the slow-roll inflation (although its amplitude is much smaller than the sensitivity of LVK). The value $\alpha = 2/3$ corresponds to the inspiral phase of compact binary coalescences (CBC)~\cite{PhysRevX.6.031018}. Finally, when a SGWB arises from the overlap of uncorrelated sources with a typical time scale of $\Delta\tau$, it results in white noise (frequency-independent) strain power at the low-frequency tail $f<1/\Delta\tau$, corresponding to $\alpha=3$. This phenomenon can be observed in phase transition origins~\cite{Romero:2021kby} or an ensemble of burst-like astrophysical sources such as supernovae~\cite{supernovae_alpha_3}.
 
The CBC background is the most likely source in the LVK frequency band~\cite{PhysRevLett.120.091101}. Therefore, whenever searching for other sources, CBCs must be simultaneously fitted using this spectrum, as we also do in the following analysis sections. The power law spectrum with $\alpha = 2/3$, which is typical of the inspiral phase, provides a good approximation as long as we consider the LVK sensitivities, while it may no longer be applicable for future detectors, and thus the contributions from the merger and ringdown phases must be included. 
\end{backgroundinformation}

\subsection{Narrowband analysis}
\label{sec:narrowband_analysis}
As seen in the formulas derived in the previous subsection, it is more convenient to work in the frequency domain rather than in the time domain. Working in the time domain requires phase information, which cannot be theoretically predicted in the case of the SGWB, where random fluctuations come from all directions. Thus, especially for constructing the filter function that requires information of the SGWB signal, the frequency domain is simpler and more practical. Therefore, in real data analysis, we first construct the cross-spectral density (CSD) using the Fourier-transformed data $s_{\rm i}(f)$ as 
\begin{equation}
 C_{12}(f) \coloneqq \frac{2}{T} s_1^*(f)s_2(f') \,.
    \label{eq:CSD}
\end{equation}
It is important to note that in actual searches, the data is divided into short segments, typically $192$ seconds each. This approach allows us to handle gaps in data due to interruptions in data collection and address potential non-stationarities in the detector noise over both short and long timescales. Each segment is analyzed individually first, and then the information from each bin is combined~\cite{Renzini_2023}. Additionally, the CSD is estimated through coarse graining by averaging over neighboring frequencies to reduce fluctuations in the PSD estimates~\cite{PhysRevResearch.3.043049}.

In practical terms, measured data are discretely sampled and the analysis is first performed in discrete bins of frequency. This approach is called \textit{narrowband analysis}. The narrowband definition of the estimator can be obtained from Eq.~(\ref{eq:cross_correlation_estimator_2}) by reducing the bandwidth of the analysis to $\delta f$
\begin{equation}
     \hat{Y}_f = \delta f \cdot 2 {\rm Re} [s_{1,f}^*s_{2,f}]Q_{f} = \delta f \cdot T ~{\rm Re} [C_{12,f}]Q_{f}\,.
    \label{eq:narrowband_1}
\end{equation}
Here ``Re'' stands for the real part of a complex number, which arises due to our focus on positive frequencies, and comes from utilizing the relation $\int_{-\infty}^\infty \dd f  ~ s_1(f)s_2^*(f) = 2 \int_0^\infty \dd f ~ {\rm Re} [s_1(f)s_2^*(f)]$. The subscript $f$ denotes the value at a frequency bin $f$ and is used when we want to emphasize that the value is obtained from data and is discrete, while the variables with the notation $(f)$ can be calculated theoretically without data. Additionally, we use a hat to indicate that the value is constructed from the data rather than being the true value. With this notation, as derived in Eq.~(\ref{eq:Q_filter}), the optimal filter $Q_f$ is rewritten as
\begin{equation}
 Q_f = \lambda\frac{\Omega_{{\rm GW},f} \gamma_{12}(f)}{f^3P_{1,f}P_{2,f}}\,.
    \label{eq:Q_filter_alpha}
\end{equation}
By rewriting Eq.~(\ref{eq:cross_correlation_estimator_average_3}), the expectation value of $\hat{Y}_f$ is given by 
\begin{equation}
 \langle \hat{Y}_f \rangle = \frac{3H_0^2}{10\pi ^2} T ~\delta f \frac{\Omega_{{\rm GW},f}}{f^3}\gamma_{12}(f)Q_f\,.
    \label{eq:narrowband_cross_correlation}
\end{equation}
The factor 2 difference in the coefficient is again due to the fact that we work with positive frequencies.
Let us take the overall normalization constant $\lambda$ to satisfy $\langle \hat{Y}_f\rangle = \Omega_{{\rm GW},f}$. With this choice of normalization, using Eqs.~(\ref{eq:Q_filter_alpha}) and (\ref{eq:narrowband_cross_correlation}), we obtain 
\begin{equation}
\lambda^{-1} = \delta f\frac{3H_0^2}{10\pi ^2} T \frac{\Omega_{{\rm GW},f}\gamma_{12}^2(f)}{f^6P_{1,f}P_{2,f}}\,.
\label{eq:lambda_value}
\end{equation}
By substituting Eqs.~(\ref{eq:Q_filter_alpha}) and (\ref{eq:lambda_value}) to Eq.~(\ref{eq:narrowband_1}), we obtain the final form of the estimator,
\begin{equation}
    \hat{Y}_f = \frac{{\rm Re}[C_{12,f}]}{\gamma_{12}(f)S_0(f)}\,,
    \label{eq:Y_alpha}
\end{equation}
where we have defined 
\begin{equation}
S_0(f)\coloneqq\frac{3H_0^2}{10\pi^2}\frac{1}{f^3} \,.
\end{equation}
It is important to note that, by setting the normalization to ensure $\langle \hat{Y}_f\rangle = \Omega_{{\rm GW},f}$, the mean value of the estimator $\hat{Y}_f$ directly corresponds to the value of the spectral amplitude of the SGWB at a frequency $f$. From Eq.~\eqref{eq:sigma_Y_2}, we can see that the associated variance is given by
\begin{equation}
\label{eq:sigma_Y_alpha}
  \sigma_{\hat{Y}_f}^2 = \frac{1}{2T\Delta f}\frac{P_{1,f}P_{2,f}}{\gamma_{12}^2(f) S_0^2(f)}\,.
\end{equation}

Finally, we aim to utilize the information from all frequency bins, as described in Sec.~\ref{sec:detection_isotropic_GWB}, which is called  \textit{broadband analysis}. The full broadband statistics can be derived by combining the narrowband statistics of all frequency bins, 
\begin{align}
        \label{eq:Y_alpha_broadband}
  & \hat{Y} \coloneqq \frac{\sum_f H^2(f) \sigma_{\hat{Y}_f}^{-2}\hat{Y}_f}{\sum_f H^2(f) \sigma_{\hat{Y}_f}^{-2}}, \\
     \label{eq:sigma_Y_alpha_broadband}
  & \sigma_{\hat{Y}}^{-2}  \coloneqq \sum_f H^2(f) \sigma_{\hat{Y}_f}^{-2}. 
\end{align}
By weighting with the variance $\sigma_{\hat{Y}_f}^2$, this approach assigns less weight to measurements with higher noise levels at frequencies where variances are larger, and it is optimal in the sense of being an unbiased, minimal variance estimator (for further details, see Sec. 6.1 of Ref.~\cite{Romano_Cornish_2017}). Furthermore, the rescaling function $H(f)$ is applied to reweight the estimate of the spectrum, optimizing the statistic for a specific shape. This means that the analysis is tailored to seek a specific spectral shape that is theoretically prepared. For example, in the case of the power law form, Eq.~\eqref{eq:Omega_GW_assumption}, the function is given by 
\begin{equation}
    H_{{\rm ref},\alpha}(f)=\left(\frac{f}{f_{\rm ref}}\right)^\alpha \,.
\end{equation}
This function may be a set of templates to be examined using Bayesian inference to determine the most probable model, as elaborated in the subsequent subsection.

In the LVK collaboration, we use a user-friendly Python-based package for the search of an isotropic SGWB in ground-based interferometer data called \textbf{pygwb}~\cite{Renzini_2023, pygwb}. Its primary purpose is to construct the optimal estimator and variance of the SGWB, as described by Eqs.~(\ref{eq:Y_alpha_broadband}) and (\ref{eq:sigma_Y_alpha_broadband}). Additionally, it facilitates Bayesian parameter estimation to constrain SGWB models. It is worth mentioning that in practical searches, the estimators $\hat{Y}_f$ calculated for each time series segment are combined for the entire observation period before averaging across frequency bins. While we do not describe the details here, interested readers can find a detailed explanation in Sec. 3. 3 of Ref.~\cite{Renzini_2023}.

\subsection{Bayesian inference}
\label{sec:likelihood_in_isotropic_GWB_searches}

The current LVK stochastic analysis employs a hybrid approach that combines frequentist and Bayesian analysis techniques~\cite{PhysRevLett.109.171102}. Specifically, certain frequentist statistics, namely $\hat{Y}_f$ and $\sigma_{\hat{Y}_f}^2$ combined for the entire observation period, are calculated as described in the previous subsection. Subsequently, their frequency series are utilized as the fundamental input data for a Bayesian analysis to calculate posterior probability distributions of model parameters, typically to provide upper bounds on the spectral amplitude of the SGWB. The hybrid frequentist-Bayesian analysis method has been utilized over the years to establish upper bounds on GWBs with various amplitudes and spectral shapes~\cite{PhysRevLett.118.121101,PhysRevD.100.061101,PhysRevD.104.022004}. Reference~\cite{Sufficient_statistic_Andrew} has demonstrated that this hybrid approach does not lose information compared to a fully Bayesian search, which produces posterior distributions from the full time-series data rather than from the frequency series of frequentist statistics.

The Bayesian approach offers advantages over frequentist methods, as with a proper choice of the prior, the posterior distribution provides more information about the true parameter within the identified region. This frequentist-Bayesian approach allows us to conveniently explore arbitrary spectra of the SGWB and is often used for parameter estimation to provide constraints on a certain cosmological model.

\begin{backgroundinformation}{Background Information - Quick summary of Bayesian inference }
In Bayesian inference, probabilities are related to knowledge about an event (or signal). For example, when we want to infer the amplitude of a SGWB ($a$), the data ($d$) are known and it is the value of $a$ that is uncertain. The relevant probability is that the amplitude has a certain value, given the data. This probability distribution is the posterior $p(a|d)$. This is unlike in frequentist inference, where the uncertainty is intrinsic to the data, and the relevant probability is that of observing the data given a signal of amplitude $a$. This probability is known as the likelihood, represented by $p(d|a)$. If we assume that $N$ time-series data samples $d=\{d_j\}$ is the sum of the signal $a$ and white, zero-mean noise noise $n_j$, i.e. $d_j=a+n_j$, with variance $\sigma$, the likelihood function can be described as
\begin{equation}
  p(d \vert a) = \frac{1}{(2\pi)^{N/2}\sigma^N}
  \exp\left[-\frac{1}{2\sigma^2}\sum_j (d_j-a)^2\right]\,.
\end{equation}
The likelihood and the posterior distribution are related by the Bayes' theorem, given by
\begin{equation}
  p(a \vert d) = \frac{p(d \vert a) p(a)}{p(d)},
\label{eq:bayes_Theorem}
\end{equation}
where $p(a)$ is the prior probability distribution for the amplitude. The prior represents the knowledge about the range and distribution of the parameter in the model (the SGWB amplitude in this example). The normalization factor $p(d)$ is the marginalised likelihood or evidence, obtained by integrating over the model parameters, $a$ in this case, $p(d)=\int p(d|a)p(a) \dd a$. 
\end{backgroundinformation}

Utilizing the estimator and the variance in each frequency bin, $\hat{Y}_f$ and $\sigma_{\hat{Y}_f}^2$, constructed from the data, the likelihood function that is used in the LVK collaboration is~\cite{PhysRevLett.109.171102}
\begin{equation}
 p(\{\hat{Y}_f\}|\Theta)
\propto
\exp
\left[
  -\sum_f \frac{(\hat{Y}_f-Y(f|\Theta))^2}{2\sigma_{\hat{Y}_f}^2}
\right] \,,
\label{eq: SGWB _likelihood}
\end{equation}
where $Y(f|\Theta)$ represents the true value of $Y$, determined by an assumed model and dependent on a set of model parameters $\Theta$. We remind the reader that the mean value of the estimator $\hat{Y}_f$ directly corresponds to the value of the spectral amplitude of the SGWB at a frequency $f$, as mentioned in the previous section. Thus, $Y(f|\Theta)$ is sometimes written as $\Omega_{\rm GW}(f|\Theta)$.

In our searches, we always compute the Bayes factor, a measure of the relative evidence provided by the data for two competing hypotheses. This factor facilitates the comparison of how effectively each hypothesis aligns with the observed data, thereby quantifying the strength of support for one hypothesis over the other. Given two hypotheses ${\cal H}_0$ and ${\cal H}_1$, the hypothesis of having a signal and the hypothesis of only having noise in the data, the Bayes factor is defined as
\begin{equation}
    \mathcal{B} = \frac{p( d | {\cal H}_1 )}{p( d | {\cal H}_0 )},
\end{equation}
where $d$ represents the data and $p( d | {\cal H}_i )$ is the evidence of model ${\cal H}_i$. It is customary to present the logarithm of the Bayes factor. For positive values, hypothesis ${\cal H}_1$ is preferred over ${\cal H}_0$. In the case of the search for a SGWB, we typically compute the Bayes factor between the hypothesis of a signal being present and the hypothesis of only noise being present in the data. Another example is the comparison between the hypothesis of the existence of only general relativity (GR) polarization (tensor polarization) and non-GR polarizations (additional scalar/vector polarizations)~\cite{Callister:2017ocg}.
\\

\begin{backgroundinformation}{Background information - Bayesian credible interval}
While all the information from a Bayesian inference run is encoded in the posterior distribution, the posterior mean and probability intervals are concise ways of expressing the results of an analysis. A Bayesian confidence interval (CI), also referred to as a credible interval (CI), represents the degree of belief about an event. It is defined by the area under the posterior distribution between two parameter values with the range containing a particular percentage of probable values. If the data are insufficient to determine the model parameter, we can obtain a Bayesian upper limit (UL), where one end of the interval is determined by the smallest value that the parameter can take. 

Practically speaking, the CI or UL is obtained by integrating the posterior distribution on a parameter until xx\% of the area under the posterior is achieved. The value of the parameter for which this occurs is the CI or UL on the parameter at the xx\% confidence level (CL). In literature, we often find $90$\% and $95$\% CI. The $90$\% is preferred for its stability, as insufficient sampling points may lead to instability in the tail distribution. The $95$\% is chosen primarily motivated by its intuitive relationship with the standard deviation ($2\sigma$). It is recommended to have a sample size of at least $10^4$ for precise computation of the 95\% CI~\cite{Kruschke2014,2019JOSS....4.1541M}. The definition for the CI and UL is represented graphically in Fig.~\ref{fig:bayesian-interval}.

Another noteworthy aspect is that the 68\% and 95\% CIs do not correspond to the notion of 1 and 2 $\sigma$ regions in two dimensions. In this case, within the 1$\sigma$ region, the Gaussian distribution contains 39\% of the volume, not 68\%. Therefore, it is important to carefully choose the wording of ``$\sigma$'' or ``\%'' when displaying the results of parameter estimation using 2-dimensional contour plots. The default corner plot output of \textbf{pygwb} utilizes the $\sigma$ notation.

\begin{figure}[!htbp]
    \centering
    \includegraphics[width=0.49\linewidth]{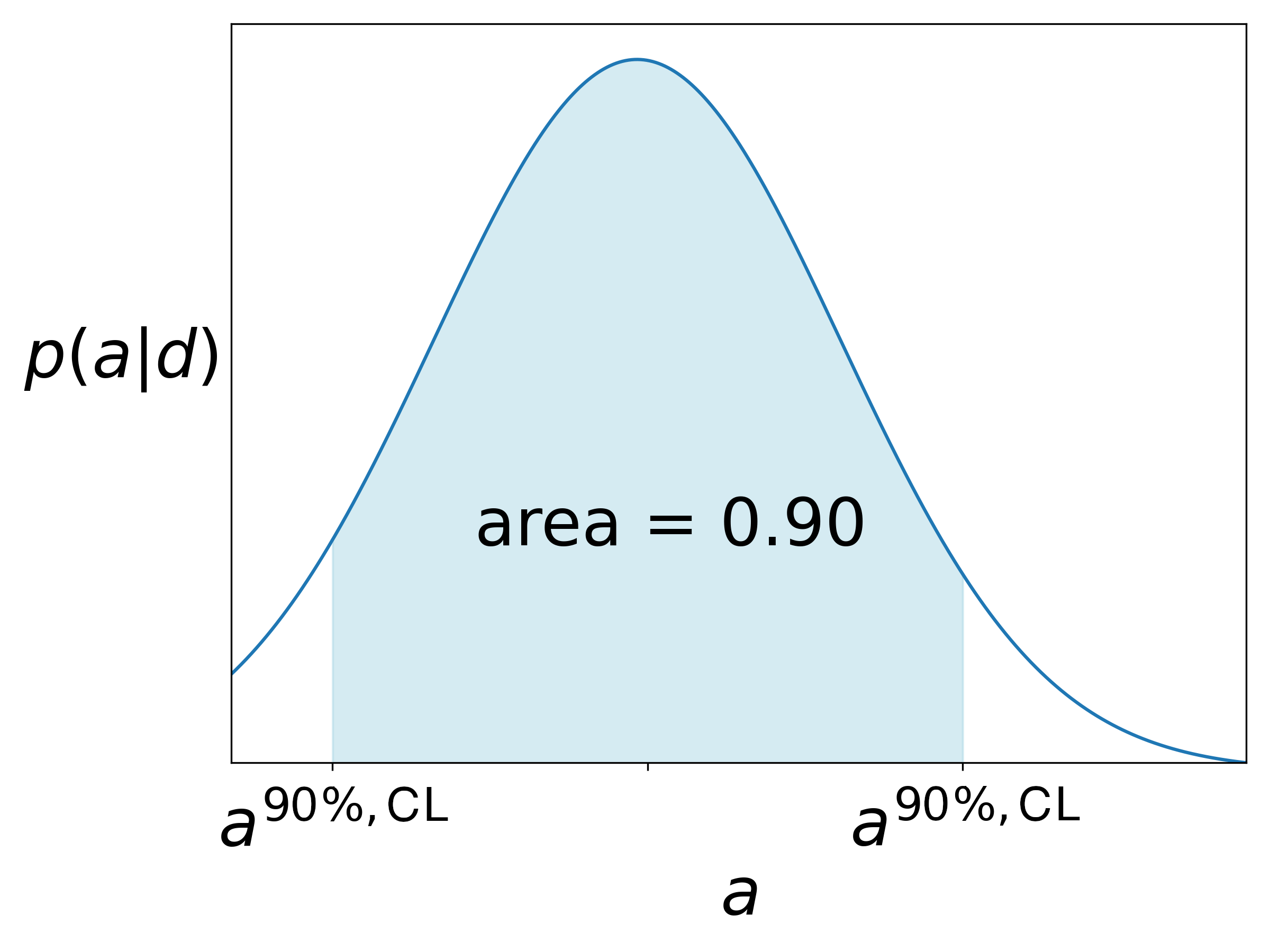}
    \includegraphics[width=0.49\linewidth]{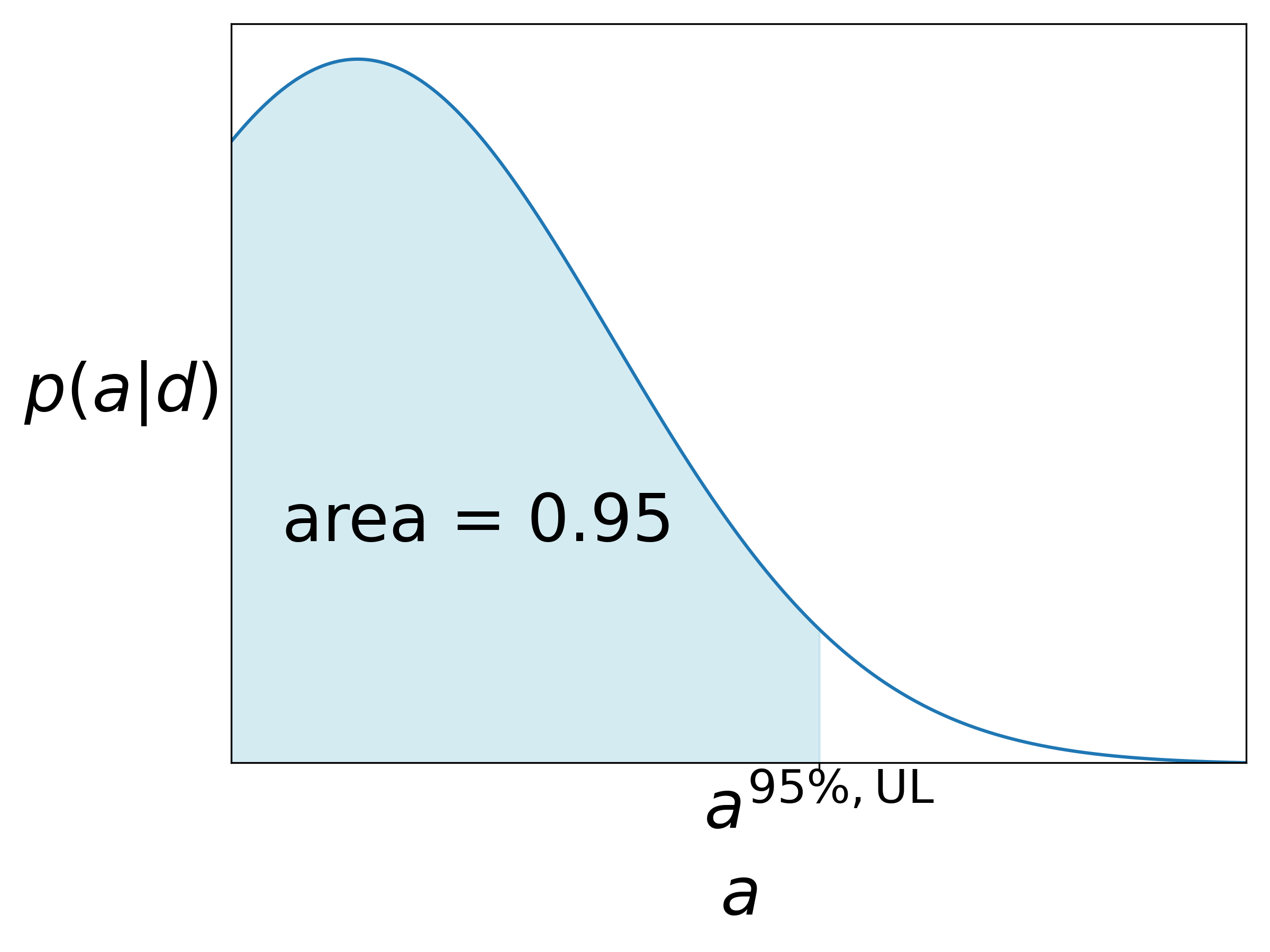}
    \caption{Graphical representations of a Bayesian 90\% credible interval (CI) and 95\% upper limit (UL), assuming that the posterior is dependent on a parameter $a$. \textbf{Left:} The area below the posterior $p(a|d)$ colored in light blue represents the CI, which coincides with 90\% of the total area. The two boundaries of the interval are commonly determined by the so-called Highest Density Interval (HDI), which is defined to contain the required area such that all points within the interval have a higher probability density than points outside the interval. In other words, the probability density at the two boundaries is the same. \textbf{Right:} The area below the posterior $p(a|d)$ colored in light blue coincides with a 95\% of the total area. The lower bound coincides with the lowest possible value $a$ can take, as determined by the prior distribution. The value of the parameter $a$ at the upper bound corresponds to the $95\%$ UL. }
    \label{fig:bayesian-interval}
\end{figure}

\end{backgroundinformation}

This section concludes with a brief discussion on priors. Prior distributions represent the beliefs or knowledge about the parameters of a model before observing the data. These distributions encode information available before data collection and, as mentioned earlier, are used in Eq.~(\ref{eq:bayes_Theorem}), to form posterior distributions after observing the data. The choice of the prior distribution is a critical aspect of Bayesian analysis, and it is essential to ensure that this choice does not unduly affect the final results.

When certain parameters are well known or constrained by either theory or experiment, their priors can incorporate this information. In contrast, non-informative or weakly informative priors are applied for parameters that are not well known. For instance, a linear or log flat prior might be used. An example of an informative prior is seen in the context of estimating astrophysical foreground amplitude in SGWB parameter estimation. Direct detections of individual CBC events help update the parameters of the CBC population, leading to better estimates of the spectral amplitude of the CBC foreground~\cite{PhysRevLett.118.121101}. As a result, more constrained priors on this parameter can be established.

\subsection{Calibration uncertainties in the search for the GWB}
The data obtained from ground-based GW detectors requires calibration to translate the digital output of the detector into a relative displacement of the test masses within the detectors. This calibration process introduces both statistical uncertainties and systematic errors,  which are commonly referred to as calibration uncertainties~\cite{PhysRevD.103.063016}. For an in-depth examination of calibration uncertainties in the Advanced-LIGO detectors, the reader is referred to Ref.~\cite{Sun_2020}.

The Bayesian analysis for a SGWB has to account for these calibration uncertainties~\cite{whelan_calib_uncertainty}. To address this, we introduce an unknown calibration factor $\Lambda$, which has a mean value of $1$ and a variance of $\varepsilon^2$. Thus the probability distribution for $\Lambda$ is given by
\begin{equation}
p(\Lambda)\propto \exp \left[-\frac{(\Lambda - 1)^2}{2 \varepsilon^2}\right] \,.
\label{eq:Lambda_distribution}
\end{equation}
Note that we impose the constraint that $\Lambda$ must be positive, and $\Lambda=1$ corresponds to perfect amplitude calibration.

The likelihood function for the Bayesian analysis, introduced in Eq.~\eqref{eq: SGWB _likelihood}, is constructed using the narrowband estimator $\hat{Y}_f$ along with its associated error $\sigma_{\hat{Y}_f}$. With the inclusion of calibration uncertainties, if the calibration is imperfect ($\Lambda \neq 1$), the set of estimators $\{\hat{Y}_f\}$ represents measurements of $\Lambda Y(f|\Theta)$, where $Y(f|\Theta)$ is a model described by parameters $\Theta$. Consequently, the likelihood is modified as follows,
\begin{equation}
 p(\{\hat{Y}_f\}|\Theta,{\Lambda})
\propto
\exp
\left[
  -\sum_f \frac{(\hat{Y}_f-{\Lambda}Y(f|\Theta))^2}{2\sigma_{\hat{Y}_f}^2}
\right] \,,
\label{eq:likelihood_calib_factor_1_baseline}
\end{equation}
and the posteriors are obtained by marginalizing over $\Lambda$, 
\begin{equation}
p(\{\hat{Y}_f\}|\Theta)
= \int \dd {\Lambda}
\, p(\{\hat{Y}_f\}|\Theta,{\Lambda}) p({\Lambda})\,.
\label{eq:posterior_by_marginalizing_over_lambda}
\end{equation}
Given our assumption of the Gaussian distribution for $\Lambda$ as stated in Eq.~\eqref{eq:Lambda_distribution}, we can analytically perform the integration, which is implemented in the \textbf{pygwb} package~\cite{Renzini_2023}.

\section{Constraints on a scalar-induced gravitational wave background}
\label{sec:constraints_SIGWB}

In this section, we present an overview of the latest constraints derived from the initial three observing runs conducted by the LVK collaboration regarding a scalar-induced GW background (SIGWB)~\cite{Kapadia:2020pnr,PhysRevLett.128.051301,Inui:2023qsd}, which is closely associated with the formation of PBH~\cite{Saito:2008jc}. This section is based on the results presented in Ref.~\cite{PhysRevLett.128.051301}. 

Among the processes for deriving constraints from LVK data, a crucial initial decision involves modeling the spectral shape of the curvature power spectrum to be explored in the analysis. Typically, an agnostic approach is adopted, and no specific inflation model is selected. It is customary to opt for a log-normal shape for the peak in the curvature power spectrum
\begin{equation}
  \mathcal{P}_\zeta(k) = \frac{A}{\sqrt{2\pi} \Delta} \exp\left[- \frac{\ln^2(k/k_*)}{2\Delta^2} \right] \,,
\label{eq:LN_peak}
\end{equation}
where the peak in the curvature power spectrum is defined by its position $k_*$ with its width controlled by the parameter $\Delta$ and its amplitude characterized by the integrated power $A$. In the $\Delta\rightarrow 0$ limit, the spectrum reduces to a Dirac delta function $ \mathcal{P}_\zeta (k) =  A\delta( \ln(k/k_*))$. 

Given the curvature perturbation spectrum, the energy density spectrum for the SIGWB is calculated using the approximated analytical expression~\cite{PhysRevD.97.123532,Espinosa:2018eve}
\begin{eqnarray}
  \Omega_{\rm GW}(k)h^2 &=& 1.62\times 10^{-5}
  \left(\frac{\Omega_{\rm R,0}h^2}{4.18\times10^{-5}}\right)
  \left(\frac{g_*}{106.75}\right)
  \left(\frac{g_{\rm *,s}}{106.75}\right)^{-4/3}
  \nonumber\\
  &\times&\frac{1}{12}\int_{-1}^1\dd x\int_{1}^\infty \dd y ~ \mathcal{P}\Big(k\frac{y-x}{2}\Big) \mathcal{P}\Big(k\frac{x+y}{2}\Big)F(x,y),
\label{eq:omega_GW_PBH}
\end{eqnarray}
where $\Omega_{\rm R,0}$ is the present value of the energy density fraction of radiation, $h$ is the dimensionless Hubble constant, and $g_*$ and $g_{\rm *,s}$ are the effective number of degrees of freedom for energy density and entropy density, respectively. The function $F(x,y)$ is given by
\begin{align}
  F(x,y)&=  \frac{288(x^2+y^2-6)^2(x^2-1)^2(y^2-1)^2}{(x-y)^8(x+y)^8}\nonumber \\&  \times \Big[\Big(x^2-y^2+\frac{x^2+y^2-6}{2}\ln\Big|\frac{y^2-3}{x^2-3}\Big|\Big)^2+\frac{\pi^2}{4}(x^2+y^2-6)^2\theta(y-\sqrt{3})\Big].
    \label{eq:PBH_F_x_y}
\end{align}
The LIGO-Virgo network exhibits its sensitivity in the frequency range between $10-500$Hz, corresponding to wavenumbers between approximately $10^{16}$ and $10^{17}$Mpc$^{-1}$. These scales re-entered the horizon when temperatures were above $10^8$GeV, so we can adopt $g_* = g_{*, {\rm s}} = 106.75$ within the Standard Model paradigm.  

\begin{figure}[!htbp]
    \centering
    \includegraphics[width=0.7\linewidth]{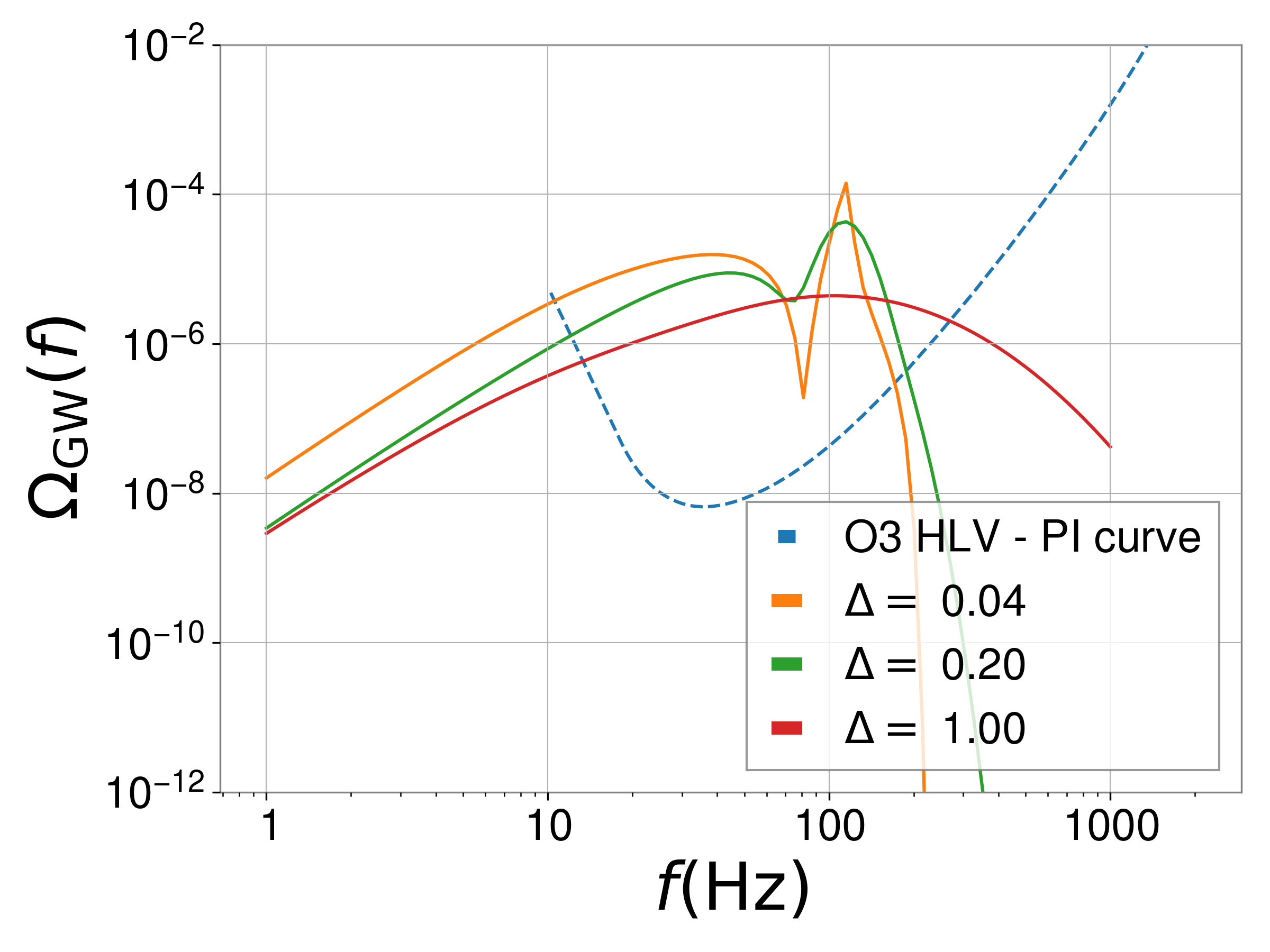}
    \caption{Example of the SGWB spectrum. Different curves correspond to different values of the width parameter $\Delta=0.04, 0.2, 1$. The other parameters are fixed as $A=1$ and $f_*=c k_*/(2\pi)=100$Hz. Additionally, we plot the O3 HLV PI sensitivity curve as a dashed blue line.}
    \label{fig:PBH_Omega_GW}
\end{figure}

Figure~\ref{fig:PBH_Omega_GW} illustrates the GWB spectrum assuming a log-normal curvature power spectrum with different values of the width parameter $\Delta$, while keeping $A$ and $f_*$ fixed. It can be observed that, for larger $\Delta$, the peak is smoothed out, and the spectral amplitude decreases. The integrated amplitude $A$ serves as the normalization of the SGWB amplitude, and the entire spectrum scales as $\Omega_{\rm GW}(f)\propto A^2$. The peak scale $k_*=2\pi f_*/c$ determines the frequency at which the SGWB spectrum peaks. The spectrum exhibits a peak at approximately the same wavenumber as the curvature power spectrum. The entire spectrum is shifted to lower or higher frequencies without changing the spectrum shape when a different value of $k_*$ is used. 

To derive constraints on the SIGWB from LVK data, we conduct a Bayesian search using the methodology outlined in Section~\ref{sec:likelihood_in_isotropic_GWB_searches}. Specifically, we utilize data products obtained from the stochastic isotropic analysis of the initial three observational runs of the LVK network, employing the former stochastic pipeline (similar to \textbf{pygwb}).  In addition to the SIGWB signal, we include the CBC foreground, $\Omega_{\rm GW,CBC}(f)=\Omega_{\rm ref}(f/f_{\rm ref})^{2/3}$ with $f_{\rm ref}=25$Hz, in the analysis, and estimate the background amplitude parameter $\Omega_{\rm ref}$ simultaneously. Therefore, the Bayesian search involves four parameters $\Theta=(\Omega_{\text{ref}}, A, k_*, \Delta)$, with their respective priors described in Table~\ref{tab:PBH_priors}. The priors on the integrated power $A$ and the peak wavenumber $k_*$ are selected so that the resulting signals could be detectable by the LVK network. Furthermore, the width parameter $\Delta$ is strategically chosen to encompass both narrow and wide peaks. Additionally, the prior on $\Omega_{\rm ref}$ is informed by prior estimates of the CBC background~\cite{PhysRevLett.120.091101}.

\begin{table}[!htbp]
\begin{center}
\begin{tabular}{|c| c|} \hline
    {Parameter} & {Prior}\\
    \hline
    $\Omega_{\rm  ref}$ & LogUniform($10^{-10}$, $10^{-7}$)\\
    $A$& LogUniform($10^{-3}$, $10^{0.5}$)\\
    $k_*/$\rm Mpc$^{-1}$ & LogUniform($10^{13}$, $10^{21}$)\\
    $\Delta$ & LogUniform($0.05$, $5$)\\
    \hline
  \end{tabular}
  \end{center}
  \caption{Summary of prior distributions used in the Bayesian analysis. We have four free parameters: $\Omega_{\rm ref}$ is the amplitude of the CBC background at $25$\,Hz, $A$ is the integrated power of the log-normal distribution of the curvature power spectrum, $k_*$ is the peak position, and $\Delta$ is the width.}
  \label{tab:PBH_priors}
\end{table}

\begin{figure}[!htbp]
    \centering
    \includegraphics[width=0.9\linewidth]{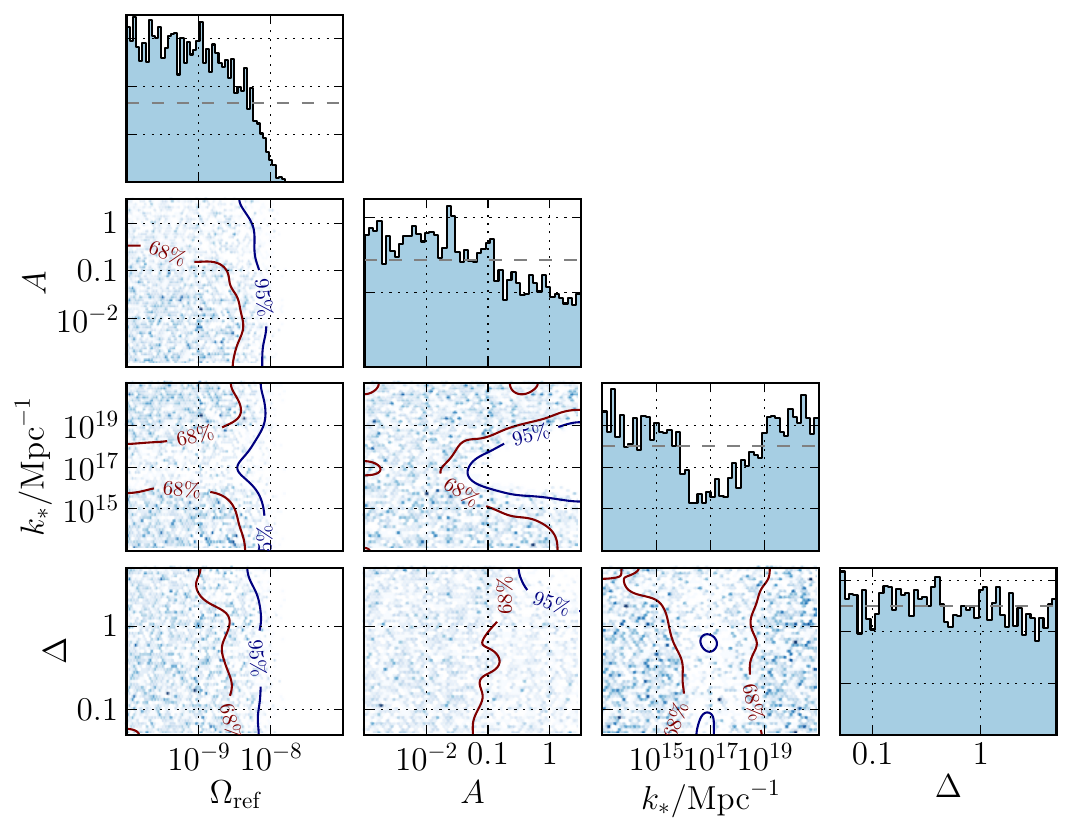}
    \caption{Figure retrieved from Ref.~\cite{PhysRevLett.128.051301}. Posterior distributions on the parameters of the curvature power spectrum describing the peak of the curvature perturbations.}
    \label{fig:PBH_corner_plot}
\end{figure}

The posterior distributions for the model parameters obtained from the Bayesian analysis are shown in Fig.~\ref{fig:PBH_corner_plot}. The posterior on $\Omega_{\rm ref}$ allows to obtain an UL of $6.0\times 10^{-9}$ at 95\% CL, which is consistent with the UL obtained in the O3 agnostic isotropic GWB search~\cite{PhysRevD.104.022004}. Data excludes a part of the parameter space in $k_*$ and $A$. Exclusions at 95\% CL are obtained in the region where the LIGO-Virgo interferometers have the highest sensitivity, $k_*\in[10^{16},10^{17}] \rm Mpc^{-1}$. The posterior of $\Delta$ shows no preference for any range of values. Finally, we have obtained a Bayes factor of $\log \mathcal{B}_{\rm noise}^{\rm CBC+PBH}=-0.8$, which indicates that there is no evidence for a GWB signal. 

The 2-dimensional contour plot as well as the 1-dimensional posterior distributions in Fig.~\ref{fig:PBH_corner_plot} are produced by \textit{marginalizing over} the other parameters. Marginalization involves integrating the posterior probability over other parameters. For example, consider that we have the the joint posterior distribution $p(\theta_1,\theta_2|d)$ and we are interested in the constraint on parameter $\theta_1$. Then the marginalized distribution of $\theta_1$ is obtained by $p(\theta_1|d)=\int p(\theta_1, \theta_2|d) ~\dd\theta_2$, and the integration range is determined by the prior range of $\theta_2$. 

It is important to note that such marginalized constraints are \textit{prior dependent}, especially when the parameters lack precise determination and the posterior distribution fails to converge to zero at the edge of the priors. For instance, the integrated power $A$ remains unconstrained for excessively large and small $k_*$ values due to the limited frequency coverage of the LVK sensitivity. Consequently, the marginalized constraint on $A$ weakens with a larger prior range of $k_*$ because we assign a greater weight to such unconstrained regions by expanding the integration range of $k_*$. Therefore, instead of discussing the marginalized constraints, we rerun the Bayesian search by fixing the values of $\Delta$ and $k_*$ to obtain ULs on $A$, thereby avoiding the influence of prior choice. The results for different combinations of $\Delta$ and $k_*$ are summarized in Table~\ref{tab:PBH_extra_runs}. The most stringent ULs are obtained for $k_*=10^{17}\rm Mpc^{-1}$, corresponding to $\sim 100$Hz, where the interferometers have their highest sensitivity. For large widths, $\Delta\gg 1$, the spectrum becomes flat and the UL on $A$ becomes independent of the scale. The most stringent UL is obtained for a narrow peak $\Delta=0.05$ and $k_*=10^{17} \rm Mpc^{-1}$. In all these runs, the ULs at 95\% CL on $\Omega_{\rm ref}$ are between $5.5\times10^{-9}$ and $6.6\times10^{-9}$, which is consistent with the isotropic results from the first three observing runs. 

\begin{table}
\begin{center}
\begin{tabular}{|c|c|c|c|} \hline
    & $k_*=10^{15}\, \rm{Mpc}^{-1}$ & $k_*=10^{17}\, \rm{Mpc}^{-1}$ & $k_*=10^{19}\, \rm{Mpc}^{-1}$ \\
    \hline
    $\Delta=0.05$ & $2.1$ & $0.02$ & $1.4$ \\
    $\Delta=0.2$ & $2.2$ & $0.03$ & $1.6$ \\
    $\Delta=1$ & $1.6$ & $0.05$ & $1.8$ \\
    $\Delta=5$ & $0.2$ &  $0.2$ & $0.3$ \\
    \hline
     \end{tabular}
  \end{center}
  \caption{Upper bounds on the integrated power $A$ of the curvature power spectrum at 95\% CL for fixed values of the peak position $k_*$ and width $\Delta$.}
  \label{tab:PBH_extra_runs}
\end{table}

\begin{figure}[!htbp]
    \includegraphics[width=0.49\linewidth]{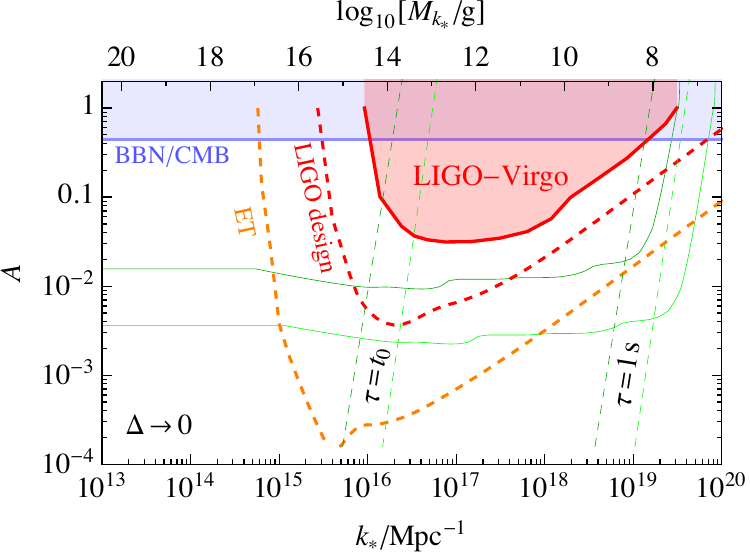}
    \includegraphics[width=0.49\linewidth]{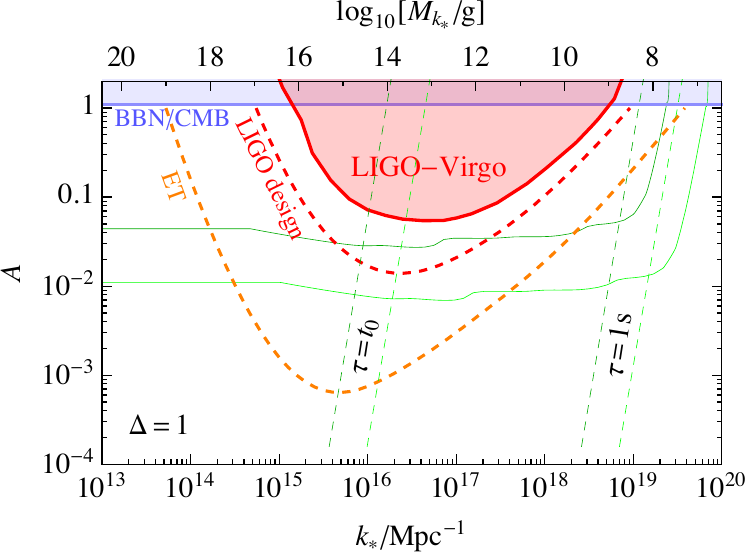}
    \caption{Figures retrieved from Ref.~\cite{PhysRevLett.128.051301}. The bounds set on $A$ from the LVK O3 data (shaded red region) are compared with other observational constraints (\textbf{Left:} for $\Delta\rightarrow 0$. \textbf{Right:} for $\Delta=1$.). The bottom and top horizontal axis represent the peak scale of the curvature perturbation $k_*$ and the PBH masses associated with the scale, respectively. The shaded blue region are indirect bounds by BBN/CMB on the SGWB abundance. The bounds from PBH formation are calculated for two benchmark cases: $\{\kappa=3.0,\delta_{\rm c}=0.65\}$ (solid dark green) and $\{\kappa=11.0,\delta_{\rm c}=0.45\}$ (solid light green). Further details can be found in the main text. The evaporation timescales of PBHs ($\tau= t_0$ and $1$s) are also indicated with dashed green lines for reference, where $t_0$ is the present time. Lastly, the dashed red line represents the bound expected from LIGO-design sensitivity, while the dashed orange line denotes that of ET. }
    \label{fig:PBH_constrains_1}
\end{figure}

In Fig.~\ref{fig:PBH_constrains_1}, the LVK bounds on the curvature power spectrum amplitude (shaded red area) are compared with other cosmological constraints for a very narrow peak ($\Delta\rightarrow 0$) in the left panel and for a relatively wide spectrum ($\Delta=1$) in the right panel. LVK bounds exhibit heightened stringency compared to indirect constraints from BBN/CMB (shaded blue area) within the $10^{16}-10^{19} ~\rm Mpc^{-1}$ region. Despite this, the PBH formation constraints (green solid lines) emerge as the most stringent across all scales. It is noteworthy, however, that with the anticipated sensitivities of Advanced-LIGO (depicted by the dashed red line) and ET (depicted by the dashed orange line), these bounds are expected to surpass even those derived from PBH formation. 

\begin{backgroundinformation}{Background information - Relation between PBH mass and scale of curvature perturbations}
PBHs are produced by critical collapse when large curvature perturbations of scale $k$ re-entered the horizon. The mass of the PBH is determined by the horizon mass, $M_k$, which is given by~\cite{Carr:1975qj}
\begin{equation}
\label{eq:M_k}
M_k \approx 1.4 \times 10^{13} M_\odot  \left(\frac{k}{\rm Mpc^{-1}} \right)^{-2}\, \left(\frac{g_*}{106.75}\right)^{1/2}
  \left(\frac{g_{\rm *,s}}{106.75}\right)^{-2/3}. 
\end{equation}
The PBH mass $M_{\rm PBH}$ then follows the scaling law~\cite{PhysRevLett.126.051303}
\begin{equation}
\label{eq:Mcrit}
 M_{\rm PBH} = \kappa M_k \left(\delta_{\rm m}  - \delta_{\rm c} \right)^{\gamma} \,,
\end{equation}
where $\delta_{\rm m}$ is the density contrast, which is related to the curvature perturbations $\delta_\zeta$ by $\delta_{\rm m}=\delta_\zeta-3\delta_\zeta^2/8$. The value of the parameter $\kappa$ depends on the procedure used to smooth the primordial perturbations~\cite{YoungS}. The value $\gamma = 0.36$ is the universal critical exponent during radiation domination~\cite{PhysRevLett.70.9}. Finally, to explain the critical collapse, it is customary to introduce the density contrast induced by curvature fluctuations at some scale, denoted by $\delta$. If the density contrast exceeds the critical value $\delta_{\rm c}$, the region collapses gravitationally and forms a PBH after the fluctuation re-enters the horizon. An estimate of the value of $\delta_{\rm c}$ was first obtained by Carr in Ref.~\cite{Carr:1975qj}, where the Jeans mass of the fluctuation was studied. This study led to $\delta_{\rm c} = 1/3$. More recent numerical studies have shown that the value of $\delta_{\rm c}$ is a bit higher. 

For the two benchmark cases to produce the green curves in Fig.~\ref{fig:PBH_constrains_1}, the critical value for the density contrast is chosen to be $0.45$ and $0.65$, which are paired with $\kappa=11.0$ and $\kappa=3.0$, respectively. The difference in the estimate of PBH abundance between the two benchmark cases reflects the uncertainties in the calculation of the PBH formation. To calculate the bounds of $A$ obtained from constraints in PBH formation, the PBH abundance from the peak in the curvature power spectrum was calculated following the procedure in Ref.~\cite{Young:2019yug}. 
\end{backgroundinformation}

\begin{backgroundinformation}{Background information - Other observational constraints}
Here we provide brief descriptions of other observational constraints shown in Fig.~\ref{fig:PBH_constrains_1}: 
\begin{itemize}
    \item {\it BBN/CMB indirect SGWB constraints: }
    BBN and CMB can indirectly constrain the amplitude of the SGWB. In the case of BBN, these indirect bounds emerge from the fact that introducing an additional radiation energy component, including the SGWB, influences the expansion rate of the Universe. This impact on the cosmic expansion can be examined through the analysis of light element abundances produced during the BBN epoch~\cite{Maggiore:1999vm}.  Similarly, the presence of a SGWB at the time of CMB decoupling introduces alterations to both the observed CMB and matter power spectra in a manner identical to massless neutrinos~\cite{Smith:2006nka}. Recent joint CMB+BBN analysis implies that $\Delta N_{\rm eff} < 0.136 $ at $2 \sigma$~\cite{Yeh:2022heq}, which leads to $\Omega_{\rm GW}h^2<1.53\times 10^{-6}$. This provides mild constraints on the curvature perturbation spectral amplitude applicable in a wide range of frequencies $f>2\times 10^{-11}$Hz, which corresponds to $M_{\rm PBH}< 10^{38}$g.
    
    \vskip.5\baselineskip
    
    \item {\it Constraints on evaporating PBHs: }
    When the PBH mass is below $5 \times 10^{14}$g, PBHs evaporate through Hawking radiation by today and cannot explain the abundance of dark matter in the present cosmological epoch. Particularly, in the PBH mass range of $10^{14}$ -- $10^{15}$g, the abundance of PBHs strongly radiating in the present Universe is strictly constrained by extragalactic $\gamma$-ray background observations. Furthermore, the PBH mass range of $10^{13}$ -- $10^{14}$g is constrained by the CMB, as PBH evaporations affect CMB observations in various ways.  Additionally, lower PBH mass ranges, $10^{9}$ -- $10^{13}$g, are constrained by BBN, as extra radiation from evaporating PBHs alter light element abundances. In Fig.~\ref{fig:PBH_constrains_1}, we illustrate the collective envelope of these constraints~\cite{Carr:2009jm,Carr:2020gox} with green curves. For reference, we also show green dashed lines labeled as $\tau = t_0$ (indicating PBHs that evaporate today) and $\tau = 1$s (representing those that completely evaporated before BBN), between which constraints of the CMB and BBN should lie.
\end{itemize}

\end{backgroundinformation}

In summary, the Bayesian analysis does not show evidence for a GWB signal, though it does allow the exclusion of certain regions within the model's parameter space. Constraints have been established based on the peak's width, its position, and the integrated power, surpassing the constraints from BBN and CMB observations in the range of approximately $10^{15}< k /\rm Mpc^{-1} < 10^{18}$. These constraints, achieving $A\simeq 0.02$ for a narrow peak centered at $k \simeq 10^{17} \rm Mpc^{-1}$, remain formidable, rivaling the restrictions arising from the PBH abundance, which must not exceed various bounds set by cosmological observations. However, we find that current ground-based experiments at their design performance and the third-generation experiments, such as ET, will reach the required sensitivity to set even more stringent upper bounds, providing a very powerful probe of the standard formation mechanism of very light PBHs. 

Finally, we note that, although we have been discussing the constraints assuming a Gaussian distribution of the primordial curvature perturbations, many inflation models predicting PBHs exhibit non-Gaussian distributions~\cite{Garcia-Bellido:2016dkw,Garcia-Bellido:2017aan,Cai:2018dkf,Atal:2019erb}, which changes the SGWB constraints. As studied in the literature~\cite{Cai:2018dig,Unal:2018yaa,Yuan:2020iwf,Adshead:2021hnm,Abe:2022xur,Perna:2024ehx}, non-Gaussianity enhances or suppresses the SGWB spectral amplitude. In particular, when we consider the quadratic local-type non-Gaussianity often parameterized by the $F_{\rm NL}$ parameter, the overall spectral amplitude is enhanced, with a more pronounced elevation at high frequencies. This leads to stronger constraints on the curvature perturbation amplitude $A$ for the large non-Gaussian case. Reference~\cite{Inui:2023qsd} discussed constraints using LVK O3 data assuming local-type non-Gaussianity with the $F_{\rm NL}$ parameterization and found that there is a strong degeneracy in the parameters of $A$ and $F_{\rm NL}$. In the limit of large non-Gaussianity ($F_{\rm NL} \gtrsim 4$), a 95\% UL is found to be $F_{\rm NL} A \leq 0.115, 0.106, 0.112$ at fixed scales of $10^{16}, 10^{16.5}, 10^{17} ~\text{Mpc}^{-1}$, respectively, for $\Delta \rightarrow 0$.

\section{Constraints on SGWB from PBH binaries}
\label{sec:constraints_binaries}
In this section, we review the constraints on a SGWB from PBH binaries. PBHs have sufficient time to form binary systems after their creation and eventually emit strong GWs when they merge. Those that occur near us can be detected as individual events, while those at high redshift cannot be resolved individually; instead, the ensemble of them can be detected as a SGWB. 

In contrast to the SIGWB described in the previous section, the theoretical prediction for the amplitude of the SGWB formed from PBH binaries is less robust due to several uncertainties. These uncertainties stem from factors such as the binary formation mechanism, the probability of disruption by a third body, and additional waveform-altering factors like eccentricity, spin, and precession, etc. (we refer the reader to Ref.~\cite{LISACosmologyWorkingGroup:2023njw} for an overview). Consequently, even if constraints on the SGWB amplitude are obtained from data, translating them to PBH parameters involves a lot of theoretical uncertainties.  Therefore, it is essential to approach the results with caution and carefully consider the assumptions that are made. In this section, we describe the most basic calculation method and briefly review the constraints obtained using the LVK data in the literature.

The energy density of the SGWB generated by PBH binaris can be derived by integrating single contributions over redshift $z$ and the two binary component masses, $m_1$ and $m_2$, as follows~\cite{Phinney:2001di}
\begin{eqnarray}
	\label{eq:Omega_z}
	\Omega_{\rm GW} (f) =  \frac{f}{\rho_{c,0}} 
    \int \dd z' \int \dd \ln m_1 \int \dd \ln m_2 \,\frac{1}{(1+z')H(z')} \nonumber \\
    \times \frac{\dd^2 \tau_{\rm merg}(z',\,m_1,\,m_2)}{\dd \ln m_1 \dd \ln m_2}  \frac{\dd E_\text{\tiny GW} (f_s)}{\dd f_s},
\end{eqnarray}
where $f_s =f (1+z)$ is the redshifted source frame frequency, with $z$ being redshift, $H(z)$ is the Hubble expansion rate, and $\tau_{\rm merg}$ is the differential merger rate per unit time, comoving volume, and mass interval. 

In the non-spinning limit, the single source energy spectrum is given by the following phenomenological expression~\cite{Ajith:2009bn} 
\begin{equation}
	\frac{\dd E_{\rm GW} (f)}{\dd f} =  \frac{(G\pi)^{2/3}}{3} \mathcal{M}_c^{5/3} f^{-1/3} \times
	\left \{ \begin{array}{rl}
		& \left( 1+ \alpha_2 u^2\right)^2 \qquad \qquad \quad  \text{for} \quad f < f_1,  \\
		& w_1 f \left( 1+ \epsilon_1 u + \epsilon_2 u^2 \right)^2 \quad  \text{for} \quad f_1 \leq f < f_2, \\
		& w_2 f^{7/3} \frac{ f_4^4}{(4 (f- f_2)^2 + f_4^2)^2} \quad \ \ \ \text{for}  \quad f_2 \leq f < f_3 \,, 
	\end{array}
	\right.
 \label{eq:dEdf}
\end{equation}
where $\mathcal{M}_c = (m_1 m_2)^{3/5}/(m_1 + m_2)^{1/5}$ is the chirp mass and $u = c^{-1}[\pi (m_1+m_2) G f ]^{1/3}$. The frequencies with index $i = 1,2,3,4$ are given by $f_i = c^3 u_{i}^3[\pi (m_1+m_2) G ]^{-1}$ with
\begin{align}
	&u_1^3  = 0.066+0.6437\eta-0.05822\eta^2-7.092\eta^3,  \nonumber \\
	&u_2^3  = 0.185+0.1469\eta-0.0249\eta^2+2.325\eta^3,  \nonumber \\
	& u_3^3 = 0.3236 -0.1331\eta -0.2714\eta^2 +4.922\eta^3,\nonumber\\
	&u_4^3 = 0.0925 -0.4098\eta +1.829\eta^2-2.87\eta^3 \,,
\end{align}
where $\eta = m_1 m_2/(m_1+m_2)^2$ is the symmetric mass ratio. The other fitting parameters are given by $\alpha_2 = -323/224 +  451\eta/168$, $\epsilon_1 = -1.8897$, $\epsilon_2 = 1.6557$, and 
\begin{align}
	&w_1  = f_1^{-1} \frac{[1 + \alpha_2 u_1^2]^2}{[1+ \epsilon_1 u_1 +\epsilon_2 u_1^2]^2},  \nonumber \\
	&w_2  = w_1 f_2^{-4/3} [1+ \epsilon_1 u_2 +\epsilon_2 u_2^2]^2 \,.
\end{align}

The merger rate is a key factor to see the model dependence of PBHs. It depends on the population of black holes and the binary formation scenario, and their details affect the spectral shape. There are two major binary formation channels~\cite{Bagui:2021dqi,Raidal:2017mfl}. One is the so-called early binary formation, where a binary is formed by a pair of closely spaced PBHs with the surrounding third object acting on the pair via a tidal force that generates the angular momentum of the binary~\cite{Nakamura:1997sm,Sasaki:2016jop}. In this case, the merger rate is given by
\begin{eqnarray}
&&\frac{\dd^2\tau_{\rm merg}(z,\,m_1,\,m_2)}{\dd\ln m_1\,\dd\ln m_2} 
=
3.4\times 10^6 
f_{\rm sup}(m_1, m_2, f_{\rm PBH})
f_{\rm PBH}^{53/37} 
\psi(m_1)\psi(m_2)
\nonumber\\
&& \quad\quad\quad \times 
\left(\frac{t}{t_0}\right)^{-34/37}
\left(\frac{m_1 + m_2}{M_\odot}\right)^{-32/37}
\left(\frac{m_1 m_2}{(m_1 + m_2)^2}\right)^{-34/37}
{\rm yr}^{-1} {\rm Gpc}^{-3} \,,
\label{eq:early_binary}
\end{eqnarray}
where $f_{\rm sup}(m_1, m_2, f_{\rm PBH})$ is a supression factor that takes into account several physical effects (see Ref.~\cite{Raidal:2018bbj,Hutsi:2020sol} for details), $f_{\rm PBH}\coloneqq \Omega_{\rm PBH}/\Omega_{\rm DM}$ is the fraction of PBHs  in dark matter, and $\psi(m)$ is the PBH mass function defined by
\begin{equation}
\psi(m)\coloneqq \frac{1}{\rho_{\rm PBH}}
\frac{\dd \rho_{\rm PBH}}{\dd \ln m} \,.
\end{equation}
The value of the suppression factor is still not clearly understood, but a plausible range is $0.001 < f_{\rm sup} < 0.1$ when considering a monochromatic PBH mass function~\cite{LISACosmologyWorkingGroup:2023njw}.

Another channel is the so-called late binary formation, where PBH binaries are formed by tidal capture in very dense halos during the matter-dominated era~\cite{Bird:2016dcv,Mandic:2016lcn,Clesse:2016ajp}. In this case, the merger rate is given by
\begin{equation}
\frac{\dd^2\tau_{\rm merg}(z,\,m_1,\,m_2)}{\dd\ln m_1\,\dd\ln m_2}=
R_{\rm clust} f_{\rm PBH}^2 \frac{(m_1 + m_2)^{10/7} }{(m_1 m_2)^{5/7}} \psi(m_1)\psi(m_2)
\quad {\rm yr}^{-1} {\rm Gpc}^{-3} \,,
\label{eq:late_binary}
\end{equation}
where $R_{\rm clust}$ is a scaling factor that depends on the PBH clustering properties and velocity distribution. A rough estimation, using a simplified model where the PBH density is uniformly enhanced within the halo, yields $R_{\rm clust}\sim 10^2-10^3$~\cite{LISACosmologyWorkingGroup:2023njw}. However, this estimate has not been thoroughly refined by considering factors such as dynamical heating, the merger and disruption history of PBH clusters, and the radial distribution of PBHs within a cluster. Additionally, the effects of an extended mass function are likely to be complex.

\begin{figure}[!htbp]
    \centering
    \includegraphics[width=0.49\linewidth]{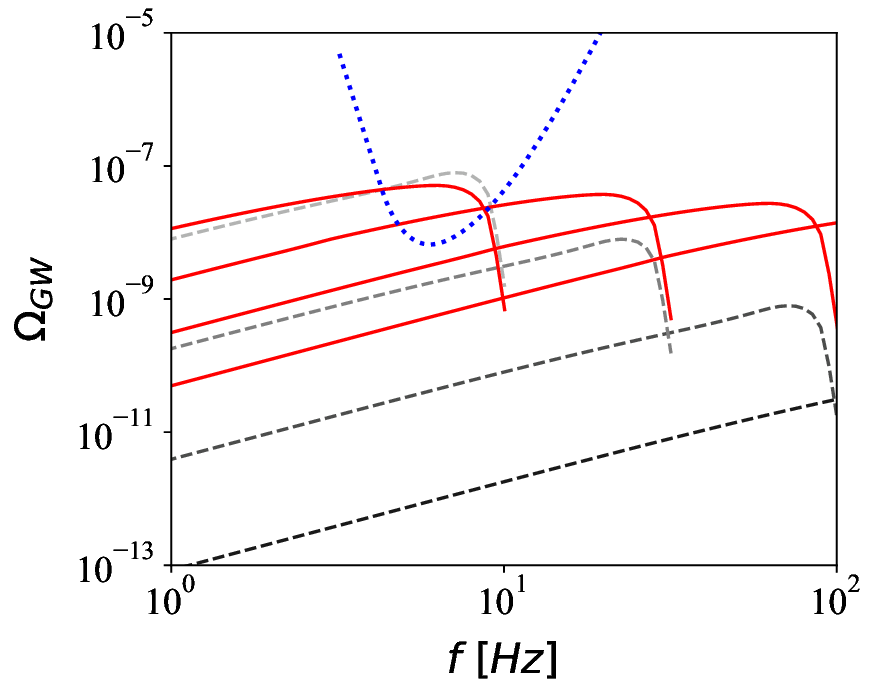}
    \includegraphics[width=0.49\linewidth]{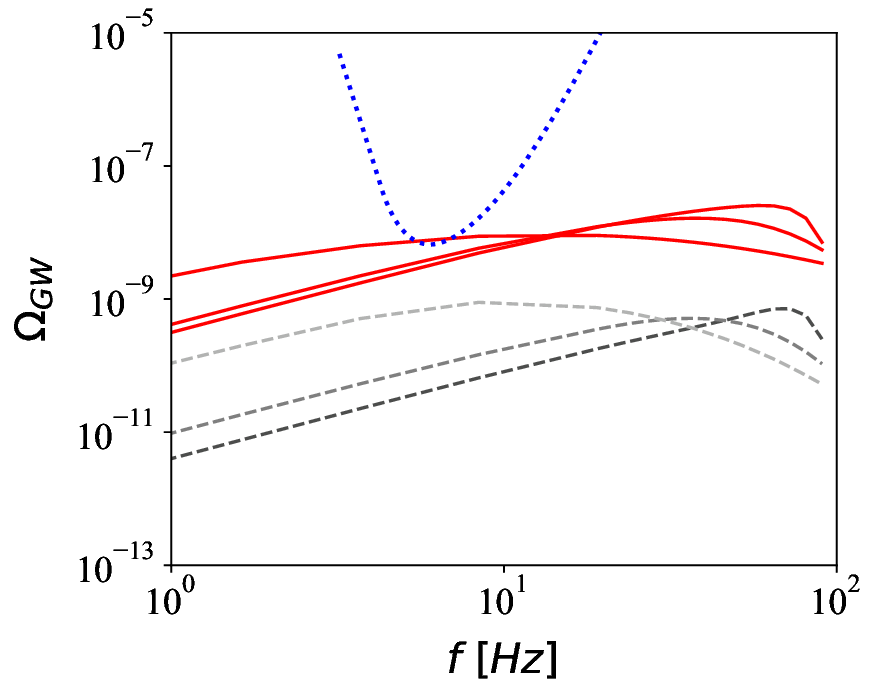}
    \caption{\textbf{Left:} Spectra of the SGWB from PBH binaries plotted assuming monochromatic mass function for different values of PBH masses (lines correspond to $M=100, 10, 1, 0.1 M_\odot$ from left to right). Solid red curves are calculated assuming early binary formation channel with $f_{\rm sup}=0.002$. Dashed black curves are calculated assuming late binary formation channel with $R_{\rm clust}=10^2$. $f_{\rm PBH}=1$ is assumed for both cases. \textbf{Right} SGWB spectra for different width of the lognormal mass function (lines correspond to $\sigma=2, 1, 0.2$ from left to right; PBH mass is fixed to be $\mu=1 M_\odot$). Solid red and dashed black curves corresponds early and late formation channels, respectively. O3 sensitivity is also plotted with the dotted blue curve for comparison.}
    \label{fig:binary_Omega_GW}
\end{figure}

Now, with predictions for the merger rate given by Eq.\eqref{eq:early_binary} or \eqref{eq:late_binary} and the energy spectrum for single sources provided by Eq.\eqref{eq:dEdf}, we can calculate the SGWB spectrum by substituting them into Eq.~\eqref{eq:Omega_z}. In the left panel of Fig.~\ref{fig:binary_Omega_GW}, we show the SGWB spectrum of PBH binaries, assuming a monochromatic mass function for different values of PBH masses and considering both early and late binary formation. We can see that SGWB from early formation channel generically dominates the one from late formation channel when the PBH mass is $M_{\rm PBH} < {\cal O}(10) M_\odot$. Roughly speaking, the redshifted PBH mass determines the peak frequency of the SGWB. The single source emits the strongest GWs at the merger, corresponding to the peak frequency $f_2$ in Eq.~\eqref{eq:dEdf}. For binary PBHs of equal mass, it is given by $f_2 = 8286 (M_\odot/M_{\rm PBH})$Hz. In the right panel, we depict the spectrum assuming a lognormal Gaussian distribution of the PBH mass function $\psi(m)=1/(\sqrt{2\pi}\sigma) \exp[-\ln(m/\mu)^2/(2\sigma^2)]$ for different width parameters $\sigma$. We observe that a wider mass function smooths out the peak of the spectrum.

As far as we consider PBH masses smaller than ${\mathcal O}(10)M_\odot$, it is most likely that we will observe the $\Omega_{\rm GW}\propto f^{2/3}$ behavior of the inspiral phase with ground-based detectors. Therefore, the astrophysical CBC background, jointly considered with the SGWB signal in the previous section, effectively represents the PBH binary background. The most conservative 95\% UL on $\Omega_{\rm ref}$ obtained in the previous section is $\Omega_{\rm ref}< 6.6\times10^{-9}$ at $25$Hz. For a monochromatic mass function, it is straightforward to show that the spectral amplitude during the inspiral phase depends on $\Omega_{\rm GW}\propto M_{\rm PBH}^{89/111}$ for early binary formation and $\propto M_{\rm PBH}^{5/3}$ for late binary formation. Therefore, the bound on $\Omega_{\rm ref}$ can provide ULs on the following combinations of parameters
\begin{eqnarray}
f_{\rm sup}f_{\rm PBH}^{53/37} (M_{\rm PBH}/M_\odot)^{89/111} \lesssim 2.6 \quad \text{for early binary formation} \\ 
R_{\rm clust} f_{\rm PBH}^2 (M_{\rm PBH}/M_\odot)^{5/3} \lesssim 200 \quad \text{for late binary formation}
\end{eqnarray}
Note that, for $M_{\rm PBH}>{\mathcal O}(10)M_\odot$, this argument is not applicable as the spectrum no longer follows the $\Omega_{\rm GW}\propto f^{2/3}$ behavior. When we have observations with better sensitivity across a wider frequency range, we must be cautious about deviations from the spectral shape of $\Omega_{\rm GW}\propto f^{2/3}$ for even smaller PBH masses, and more precise considerations are needed. 

Reference~\cite{Mukherjee:2019oma} performed parameter estimation for a SGWB from PBH binaries using LVK data. As seen in their work, with the current sensitivity of the experiments, we are unable to obtain useful constraints on the model parameters, but third-generation detectors will be a powerful tool to extract information on PBH populations~\cite{Mukherjee:2021itf}.

\section{Conclusions}
\label{sec:conclusions}

In this review, we have explored potential signals of PBHs in SGWB observations. We have delved into two distinct types of signals: those associated with PBH formation in the early stages and signals arising from the binaries of PBHs. We have discussed their current constraints, focusing on observations from the recent LVK O3 run. These two observables probe completely different epochs of the Universe; one provides access to the very early radiation-dominated phase, while the other probes PBHs in the late Universe. The scales of PBHs probed by the LVK detectors are also very different: the former imposes limits on PBHs of $10^{-24}M_\odot$ to $10^{-18} M_\odot$, while the latter exhibits sensitivity to ${\cal O}(10^{1-100}) M_\odot$.

\begin{figure}[!htbp]
    \centering
    \includegraphics[width=0.49\linewidth]{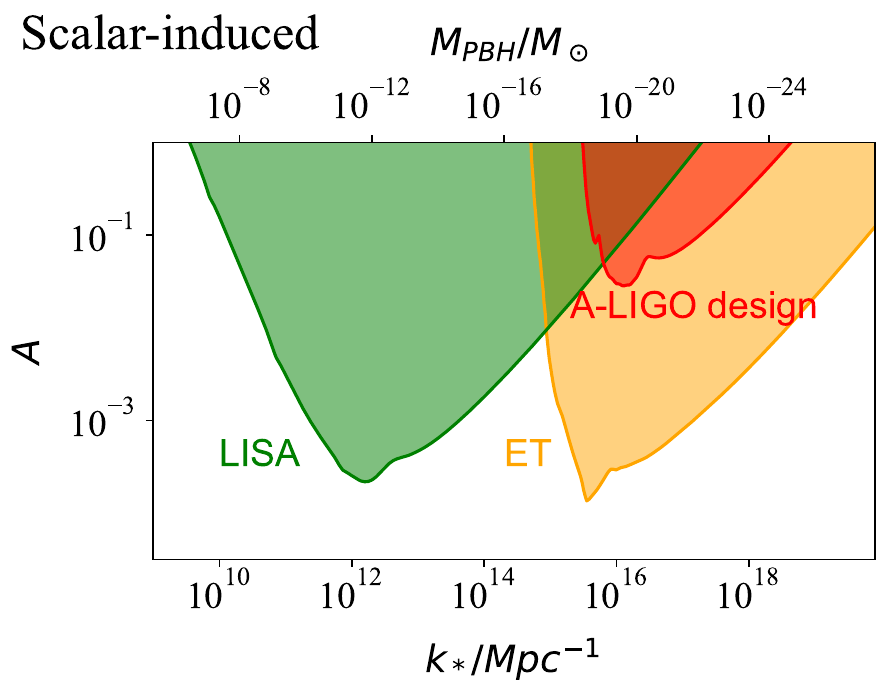}
    \includegraphics[width=0.5\linewidth]{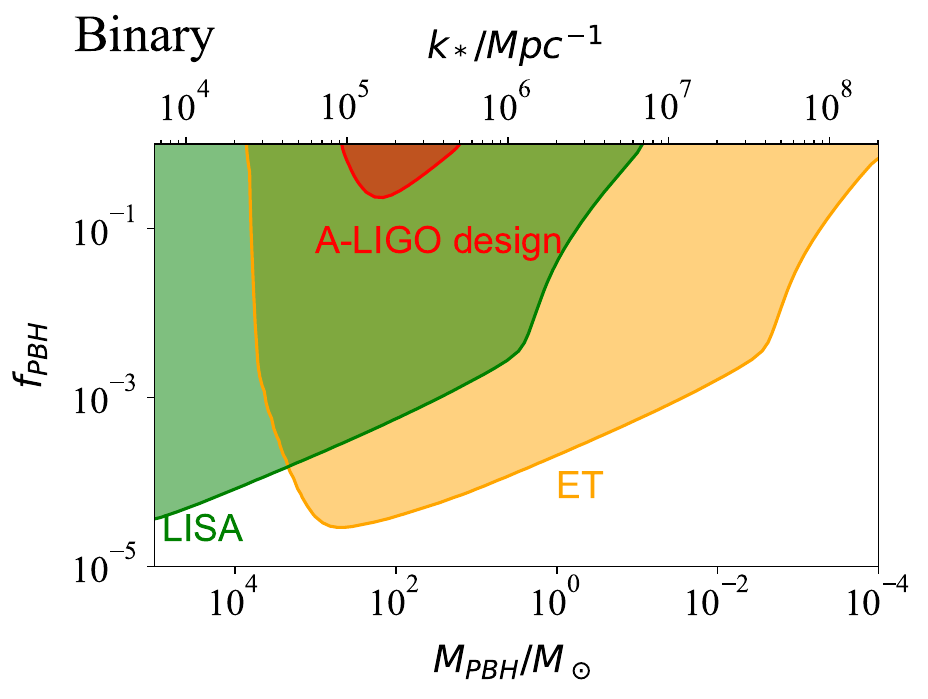}
    \caption{Accessible parameter space with future experiment sensitivities \textbf{Left:} for the spectral amplitude of the curvature power spectrum $A$, achievable through the SIGWB in the future is shown as functions of both $k_*$ and $M_{\rm PBH}$, and \textbf{Right:} for the fraction of PBH $f_{\rm PBH}$ through the PBH binary background (assuming the early binary formation channel with $f_{\rm sup}=0.002$). For both cases, monochromatic PBH mass function is assumed. The curves are obtained assuming 1 year of observation by Advanced-LIGO design sensitivity, ET, and LISA, with a conservative detection threshold of ${\rm SNR}=8$. 
}
    \label{fig:Exclusions_Ak_plane}
\end{figure}

Present constraints already provide meaningful limits, and even stronger constraints will become available with enhanced sensitivity in ongoing and upcoming observation runs, O4 and O5. Moreover, third-generation detectors, such as the Einstein Telescope (ET) and the Cosmic Explorer (CE), are expected to significantly increase sensitivity, reaching $\Omega_{\rm GW}\sim 10^{-13}$ and expanding the frequency range. Additionally, while not explored in detail here, space-based GW detectors like LISA, TianQin, Taiji, DECIGO, and other exciting proposals can offer intriguing constraints in a similar manner. Furthermore, pulsar timing arrays offer another avenue to probe SGWB at nano-Hz frequencies. In such cases, the corresponding PBH mass tends to shift toward higher masses because of the lower targeting frequencies of these experiments. All the advancements in GW probes will enable us to explore a much broader parameter space of PBH models. We conclude with Fig.~\ref{fig:Exclusions_Ak_plane}, which presents the anticipated future explorations for PBHs across a wide range of mass scales, highlighting the immense potential of future GW experiments to explore new physics.

\bibliographystyle{unsrt}
\bibliography{chapter_27}

\end{document}